\documentstyle[a4,epsf,amsmath,graphicx,colordvi,twocolumn,namedplus]{article}
\newcommand{\bbox}[1]{\pmb{#1}}
\def\bc{\begin{center}}
\def\ec{\end{center}}
\def\be{\begin{equation}}
\def\ee{\end{equation}}
\def\bea{\begin{eqnarray}}
\def\eea{\end{eqnarray}}
\def\Re{\text{Re}\>}
\def\Im{\text{Im}\>}
\def\bu{{\bf u}}
\def\bv{{\bf v}}

\def\M{{\sf M}}

\def\L{{\sf L}}

\def\A{{\sf A}}
\def\bI{{\sf I}}
\def\para{{\|}}

\def\beeta{\bbox{\eta}}
\def\bI{{\bf I}}
\def\bP{{\sf P}}

\renewcommand{\i}{{\rm i}}
\renewcommand{\d}{{\rm d}}
\newcommand{\e}{{\rm e}}

\newcommand{\cc}{{\rm c.c.}}
\newcommand{\myanot}[1]{\item}    
\renewcommand{\be}{\begin{equation}}
\renewcommand{\ee}{\end{equation}}
\def\bea{\begin{eqnarray}}
\def\eea{\end{eqnarray}}
\renewcommand{\Re}{{\rm Re}\,}
\renewcommand{\Im}{{\rm Im}\,}
\renewcommand{\bu}{{\bf u}}
\def\bv{{\bf v}}
  
     \def\M{{\sf M}} 
 \def\para{{\|}}

\def\bxi{\bbox{\xi}}
 \def\bI{{\bf I}}

\renewcommand{\i}{{\rm i}}\renewcommand{\d}{{\rm d}}
\renewcommand{\e}{{\rm e}}
\renewcommand{\cc}{{\rm c.c.}}
\title{A Learning rule for Place Fields in a Cortical Model:Theta phase precession 
as a {network} effect}
\author{Silvia Scarpetta,
Maria Marinaro\\
\small
Dept. of Physics ``E.~R.\ Caianiello''  University of Salerno,
84081 Baronissi (SA), IT\\
\small
INFN, Sezione di Napoli,   Gruppo Collegato di Salerno\\
\small
INFM, Unit\`a di Salerno; 
\small
IIASS, Vietri sul Mare, (SA) IT
}
\begin{document}\maketitle
\abstract
We show that a model of the hippocampus introduced recently by
Scarpetta,  Zhaoping \& Hertz ([2002] Neural Computation 14(10):2371-96),
 explains the theta
phase precession phenomena. 
In our model, the theta phase precession
comes out as a consequence of the   {associative-memory-like} network dynamics,
 i.e. the network's ability to imprint and recall 
oscillatory patterns, coded  both  by phases and amplitudes of oscillation.
The learning rule used to imprint the oscillatory states is a natural
generalization of that used for static patterns in the Hopfield model,
and is based on the spike time dependent synaptic plasticity (STDP),
experimentally  observed.

In agreement with experimental findings, 
the place cell's activity appears at consistently
earlier phases of subsequent cycles of the ongoing theta rhythm
during a pass through the place field,
while the oscillation amplitude of the place cell's firing rate
increases as the animal approaches the center
of the place field and decreases as the animal leaves the center.
The total phase precession of the place cell is
 lower than $360^o$, in agreement with experiments.
As the animal enters a receptive field
the place cell's activity comes slightly less than $180^o$ 
after the phase of maximal
pyramidal cell population activity, in agreement with
the findings of Skaggs et al (1996).
Our model predicts  that the theta phase 
is much  { better} correlated with
location than with time spent in the receptive field.
Finally,  in agreement with the
recent experimental findings of Zugaro et al (2005),
our model predicts that theta phase precession persists
after  transient intra-hippocampal  perturbation.

\section{Introduction}
In this paper we will show how
the model of the hippocampus introduced
in Scarpetta et al. (2002)
 is  able to explain the theta
phase precession phenomena: the
observation that oscillation phases of the  activities of
hippocampal place cells
 systematically shift
relative to the theta rhythm
 as the animal traverses the place fields (O'Keefe Recce 1993;
Skaggs et al 1996).

The first part of the paper reviews the model, its learning rule,
and the recall dynamics.
Briefly, the model's learned
states are
encoded by both the amplitudes and the phases of the oscillating
neural populations, enabling more efficient and robust information coding
than in conventional models of associative memory.
The learning rule to imprint such oscillatory states is a natural
generalization of that used for static patterns in the Hopfield model,
and is based on the STDP synaptic plasticity observed experimentally.
In particular, long-term potentiation and depression of the synaptic efficacies
depends on the relative timing of the pre- and post- synaptic activities
during learning.
A further step is introduced to allow imprinting
of correlated patterns properly.

The second part of this paper shows how the model accounts for theta phase
 precession observed in rat hippocampus.
 { During learning of each spatial location, our model imprints an oscillatory
pattern of specific amplitude and phase relationship between neurons by modifying neural
synaptic connection strengths. After learning,
the network's response to external inputs reveals the encoded phase information, and
thus phase precession, even though the external inputs do not have the phase information.
The phase precession effect comes out
from the crucial ability of our model to  store  both the amplitudes relationship
and the phases relationship of the oscillatory patterns in the synaptic connections.
Both the bell-shaped firing tuning curves (receptive fields) and
the phase precession phenomena  are a result of the
network synaptic connections.
In particular, the phase precession
is an effect of the overlap between the imprinted
place fields:  if adjacent
receptive fields do not overlap
 the firing tuning curve of
each place cell is preserved but the
phase precession does not occur.
}
 {The model  agrees
 with the observation (Skaggs et al 1996)
that the first spikes,
as the animal enters
the receptive field, come near the
phase of the minimal pyramidal population average activity.
The  total phase precession of the place cells can be
close to $360^o$, but not larger than $360^o$
as found by O'Keefe Recce (1993) and  Skaggs et al (1996).
}
  {
In  our model,
 the theta phase of the place cell is a function of
the animal location,
almost insensitive to the animal'
running speed (for speeds
that do not exceed a critical value related
to the network connections and parameters),
and therefore
 the theta phase shows a better correlation
 with the  animal
 location  than with time,
 in agreement with the experimental results
of O'Keefe Recce (1993) and  Huxter et al. (2003).
}

Numerous computational models
(Burgess et al 1994,
Tsodyks et al 1996,
 Wallenstein et al 1997, Jensen and Lisman 1996,
 Bose et al 2000, Booth et al 2001, Yamaguchi  2003,
and Lengyel et al 2003 )
have attempted to account for the mechanisms that underlie the
phase precession observation, but
 a fully-satisfactory explanation  is still lacking.
Recently, to contrast the predictions of different models
of phase precession, Zugaro et al (2005) have transiently turned off
neuronal discharges for up to 250 ms and reset the phase of theta oscillations
by stimulating the commissural pathway in rats. After recovery from silence,
phase precession continued: the theta phase of the place cell immediately after the
recovery was correlated with the new location of the animal, despite the 
transient silence and the theta phase reset.
Our model is in agreement with these recent experimental findings.

\section{The model}
The model structure and elements are
based on the physiological and anatomical findings in the CA3
 hippocampal region.
  These regions contain
the principle pyramidal cells and the inhibitory interneurons.
The pyramidal cells project long range axons to other pyramidal cells and
 interneurons, whereas the interneurons  project more  locally.
The model is based on the one introduced and analyzed
 by one of us with Li Zhaoping
and John Hertz
in a recent paper (Scarpetta et al 2002).
 We review here its main
features, more details can be found in
the paper above mentioned. The state variables, modeling the membrane potentials, are
$\bu=\{u_1, \ldots,$ $ u_n\}$ and $\bv=\{v_1, \ldots, v_n\}$
respectively for the  excitatory and  inhibitory units.
  (We denote vectors by bold  fonts.) 
The unit outputs, representing the probabilities of the
cells firing (or instantaneous firing rates) are given 
by $g_u(u_1), \ldots,\, g_u(u_n)$ and 
$g_v(v_1), \ldots,\, g_v(v_n)$,
 where $g_u$ and$g_v$ are sigmoidal activation functions that model the neuronal input-output
relations.  The equations of motion are
\bea\dot{u}_i&=& -\alpha u_i -\beta_i^{0} g_v(v_i)+ \sum_{j} J_{ij}^{0} g_u(u_j) + I_i  ,                                      
  \nonumber      
    \\
\dot{v}_i&=& -\alpha v_i +\gamma_i^{0} g_u(u_i)+ \sum_{j \neq i} W_{ij}^{0} g_u(u_j) .
        \label{eqnsv}
\eea
where 
$\alpha^{-1}$ is the membrane time constant (for simplicity assumed the same
for excitatory and inhibitory units),
$ J_{ij}^{0}$ is the synaptic strength from excitatory unit $j$
to excitatory unit $i$, $ W_{ij}^{0}$ is the synaptic
strength from excitatory unit $j$ to inhibitory unit $i$,
$\beta_i^0$ and $\gamma_i^0$ are the local inhibitory 
and excitatory connections within the
E-I pair $i$, and $I_i(t)$ is the net input from other parts of the brain.
The {model units} represent local populations of
biological neurons that share common input, so the number of
neurons represented by an excitatory unit may be different from 
the number of neurons represented by an inhibitory unit.
In this minimal model we omit inhibitory 
connections between pairs, since the real anatomical
long-range connections appear to come predominantly 
from excitatory cells.
{
The sensory input $I_i(t)$ to the system, which drives the excitatory units,
has a static part $\bar I_i$ and an oscillatory part $\delta I_i(t) $
modulated at the
theta frequency.}
 The static part $\bar \bI$ of the input determines 
a fixed point $(\bar{\bu},\bar{\bv})$,
 given by the solution of equations $\dot{\bu}=0, \dot{\bv}=0$ with $\bI = \bar \bI$.

  Linearizing the equations  (\ref{eqnsv})
 around the fixed point leads to 
  \bea\dot{u_i} &=& -\alpha u_i -\beta_i v_i + \sum_j J_{ij} u_j + \delta I_i \nonumber
  \\
  \dot{v}_i&=& -\alpha v_i +\gamma_i u_i + \sum_j W_{ij} u_j  \label{eqnslin}\eea 
  where $u_i$ and $v_i$ are now deviations from the fixed point,
  $\delta \bI \equiv \bI - \bar \bI \,$ is the oscillatory part of the sensory input,
 $\beta_i = g'_v(\bar v_i) \, \beta_i^{0}$,
  $\, \gamma_i = g'_u(\bar u_i) \, \gamma_i^{0}$,
  $\, W_{ij}=g'_u(\bar u_j) \, W_{ij}^{0}$,
  $\, J_{ij}=g'_u(\bar u_j) \, J_{ij}^{0}$.  Henceforth, for simplicity, we assume
  $\beta_i = \beta$, $\gamma_i=\gamma$, independent of $i$.

  Eliminating the $v_i$ from (\ref{eqnslin}), we have
  the  second order differential equations
  \begin{equation}
  \ddot{\bu} + (2\alpha -{\sf J}) \dot{\bu} + [\alpha^2 - \alpha {\sf J} + \beta (\gamma +{\sf W})] \bu  =
  (\partial_t + \alpha)\delta {\bf I}.  		
  \label{ddu}
  \end{equation}
  (We use sans serif to denote matrices) or, equivalently,
  \begin{equation} \left[(\partial_t +\alpha)^2 +\beta \gamma \right] \bu =\M  \bu+ (\partial_t +\alpha)\delta \bI
  \label{eqlin1}  
  \end{equation}
  \begin{equation}\mbox{where} \qquad \M=  (\partial_t +\alpha)  {\sf J}-\beta {\sf W} \, ,
  \label{MMM}
  \end{equation}      

  The terms in the square bracket describe the local E-I pair dynamics,
 while ${\sf M}$ gives an effective coupling
   between the oscillating E-I pairs.
    Learning imprints patterns into $\M$ through the long range 
connections $\sf J$ and $\sf W$.
     After learning, ${\bu}$ depends on how $\bI$ decomposes into the 
eigenvectors of $\M$.
      Thus the network can selectively amplify  or distort $\bI$ in 
an imprinted-pattern-specific 
manner and thereby functions as 
 input mapping into the
      subspace spanned by the imprinted patterns.

Following  Scarpetta et al (2002), we distinguish a learning mode, in which the
      oscillatory patterns are imprinted while the
 connections $\sf J$ and $\sf W$ are ineffective,
      from a recall mode, in which connection strengths are effective and fixed.
      With appropriate learning kernel, the network
      during recall responds strongly (resonantly) to inputs
      similar to those learned  or linear combinations
      of them, and weakly to other inputs not related to the imprinted patterns.
      The phases and the
   amplitudes of the oscillatory response will be a linear combination
      of the phases and the amplitudes, respectively, of the 
      imprinted  oscillations recalled by the input. 

      \subsection{Learning   {oscillatory patterns}}     
 {
First, let's }
 imprint a  single  oscillatory
pattern, $\bP^\mu (t) = \bxi^{\mu} e^{-\i \omega_0 t}+ \cc
$
where $\cc$ denote complex conjugate,
 {
and $\bxi^{\mu}$ is a complex vector with components
$\xi_j^\mu= |\xi_j^\mu| e^{i \phi^\mu_j }$.
       Following  Scarpetta et al (2002),
 the synaptic strength $J_{ij}$ is
       shaped by the correlations of the activities,
$u_j(t')$ and $u_i(t)$, 
in pre-synaptic and post-synaptic cells
       through
       \bea
J_{ij} &=& \frac{1}{NT}\int_0^T\d t \int_{-\infty}^{\infty}\d t^\prime\,u_i(t) A_{J}(t-t^\prime) u_j(t^\prime)
                        \label{jw1}
			\eea
			where
	 $A_J(t-t^{\prime})$ is the learning kernel that measures the
strength of the synaptic change at 
         time delays $\tau=t-t'$ between the 
         pre-synaptic
	 and
	 post-synaptic activities.
In the case when $A_J(t-t^{\prime}) = \delta (t-t')$
 the learning rule becomes the conventional Hebbian learning with
	 $J_{ij} \propto \int dt u_i(t)u_j(t)$ used in Li and Hertz (2000).
	 To model the experimental results of STDP (Markram et al 1997, Bi \& Poo 1998)
the kernel $A(\tau)$ was taken as an asymmetric function of $\tau$,
	 mainly positive (LTP) for $\tau>0$ and mainly negative (LTD) for $\tau<0$. 
	 Analogously for the synaptic strength
	 $W_{ij}$ we write
	 \bea
W_{ij} &=& \frac{1}{NT}\int_0^T\d t \int_{-\infty}^{\infty}\d t^\prime\,v_i(t) A_{W}(t-t^\prime)u_j(t^\prime).     
	             \label{jw}\eea
   Since the connections $\sf J$ and $\sf W$ are ineffective during learning,
   the responses $u_i$ and $v_i$ are proportional to the imprinting
   input $P^\mu_i (t)$.  Substituting them
   into eqs. (\ref{jw1}) and (\ref{jw}) yields connections
    \bea
J_{ij}^{\mu}  &=&  \frac{2}{N}\Re [\tilde A_{J}({\omega_0}) \,\xi_i^{\mu} \xi_j^{\mu *} ]\nonumber
    \\
    W_{ij}^{\mu}  &=& \frac{2\gamma}{N} 
     \Re
\left[
 \frac{\tilde{A_{W}}({\omega_{0}})}{\alpha -\i \omega_0} \xi_i^{\mu}\xi_j^{\mu *} 
\right] ,
     \label{JWlearning}\eea
     where
     \bea
\tilde A_{J,W}(\omega) &=& |\chi_0(\omega)|^2
     \int_{-\infty}^{\infty}\d \tau\,A_{J,W}(\tau) \e^{-\i \omega \tau}             \label{Atilde}	\\
     \mbox{with~~~}
     \chi_0(\omega)&=&\frac{\alpha -i \omega_0}{\alpha^2 +\beta\gamma - \omega_0^2 - 2 \i \omega_0\alpha }
     \eea
      Note that $\Im \tilde A_{J,W}(\omega) =0$ if $A_{J,W}(\tau )$ is
      symmetric in $\tau$ and that $\Re \tilde A_{J,W}(\omega) =0$ if
      $A_{J,W}(\tau )$ is antisymmetric.
      The dependence of the neural connections ${\sf J}$ and ${\sf W}$ and the oscillator couplings $\M$ on
      $\xi_i^{\mu} \xi_j^{\mu *}$ is just a natural
      generalization of the Hebb-Hopfield factor $\xi_i^{\mu} \xi_j^{\mu}$
      for (real) static patterns.  This becomes particularly clear 
      for kernels satisfying the following matching condition:
      \begin{equation}
      \tilde A_{J}(\omega_0)=\frac{ \beta \gamma }{ \alpha^2 + \omega_{0}^2}  \tilde A_{W}(\omega_0), ~~~~~ \mbox{at} ~~~~~\omega = \omega_0.
      \label{match_case}\end{equation}
      Then the oscillator coupling simplifies into a familiar outer-product form for
      complex vectors $\bxi$:
      \begin{equation}M_{ij}^\mu =-2 \i \omega_0 \tilde A_{J} (\omega_0 )\xi_i^{\mu} \xi_j^{\mu *}/N,   
                                                  \label{Msimple}
						  \end{equation}
 
 In the following we will always
 consider the case of matching kernels (\ref{match_case}).

We are interested in generalizing eq. (\ref{JWlearning}) and (\ref{Msimple}) to
the case in which multiple correlated patterns have to be imprinted.
   To construct the corresponding matrices for multiple patterns, 
$\bxi^\mu$, $\mu=1,..,N$, 
 we use the rule
proposed by Personnaz et al (1985) and
Diederich and Opper (1997) 
(see also Hertz el al 1991)
 for learning static correlated patterns
    without errors  in the Hopfield model. 
Thus we write, for N, linearly independent, correlated patterns $\bxi^\mu$,    
     \be J_{ij}= Re \left( \sum_{\mu\nu} \tilde A(\omega_0)
  {\xi_i^\mu} (Q^{-1})_{\mu\nu} {\xi_j^\nu}^* \right) \label{rule}
     \end{equation}
     where the $N\times N$  matrix $Q$, defined by
\be
Q_{\mu\nu} = \frac{1}{N} \sum_j {\xi_j}^{\mu *} {\xi_j}^\nu \, ,
\ee
     is the natural generalization of the correlation matrix Q used 
by Personnaz et al (1985)  and Diederich and Opper (1997)
 for the Hopfield model.
The rule (\ref{rule}) is not local
     and not iterative. However it has been proved in Diederich and Opper (1997)
 that, in the Hopfield framework, there exists
     a local rule that uses only  successive presentation of one pattern at a time that
     converges exactly to the synaptic efficiencies obtained with the Q-rule.
  {
    Work is in progress to generalize their procedure,
and obtain the same result also in our case.}
 In the following, we have used  the rule (\ref{rule}).
      With the rule (\ref{rule}),
   we imprint N oscillatory patterns $P^\mu = {\xi^\mu} e^{\i \omega_0 t}$, all
 with the same frequency $\omega_0$.
      Then the M matrix  becomes:
      \begin{equation}M_{ij} = \frac{1}{N}
 \sum_{\mu\nu}-2 \i \omega \tilde A_{J} (\omega_0)\,
\xi_i^{\mu} \, (Q^{-1})_{\mu \nu} \, \xi_j^{\nu *},   
                                                  \label{Mfinal}
						  \end{equation}

 \subsection{The recall mode}
 {
In the recall mode the network dynamics is governed by
the
matrix $M$,
 learned during the learning mode.
 }
Keeping in mind eq. (\ref{Mfinal}),  it is clear 
that the imprinted oscillations (even though not-orthogonal)
 are eigenvectors of the connection matrix M,
    \be
    \M  \,\bxi^\gamma = -2 \i \omega_0 \tilde A(\omega_0) \bxi^\gamma. \ee
   
    Thus,
  {after the transient (which is governed
by the network reaction time that depends both on the membrane time
constant of the neurons, $\alpha$, and the network synaptic connections)},
 the response $\bu$ to an input  that perfectly match one of
    the imprinted vector is
    \be\bu= \chi  \bI\ee
    with the linear response coefficient (or susceptibility)
    \be
    \chi = \frac{\alpha -i \omega_0} {\alpha^2 +\beta\gamma - \omega_0^2 - 2 \i \omega_0 \alpha + 2 \i \omega_0 \tilde A(\omega_0)}
    \label{constraints}
\ee
    A properly chosen learning kernel $A(t)$ satisfying
    \bea\alpha^2 +\beta\gamma - \omega_0^2  + 2 \i \omega_0
 \Re [\tilde A(\omega_0)] \rightarrow 0
   \nonumber
 \\
    \alpha - \Im [ \tilde A(\omega_0) ] \rightarrow 0_+
   \label{Constraints}
    \eea
    would produce a very strong susceptibility $\chi $ and thus a 
resonant response to
    match input. Meanwhile, an input pattern $\bI$
    outside the subspace of the imprinted vectors evoke a weaker response
    $\bu= \chi_0  \bI$.
 {
In general, an input $I=\bxi e^{i \omega t}$ that
overlaps with several imprinted patterns will evoke
a correspondingly mixed resonant response
\be
\bu= \chi \bI_{\para} + \chi_0  \bI_{\perp}
\label{ii}
\ee
where $ \bI \equiv  \bI_\para +  \bI_\perp $, 
 $\, \bI_{\para} = \sum_\mu \langle \beeta^{\mu} | \bxi \rangle
\bxi^{\mu} \e^{-\i \omega t} \equiv \sum_\mu N^{-1}(\sum_j \eta_j^{\mu *} \xi_j )
\bxi^{\mu} \e^{-\i \omega t}$, and $\beeta^\mu= \sum_\nu Q_{\nu \mu} \bxi^\nu$.
\\
This feature enables
the system to interpolate between imprinted patterns, and to perform
an elementary form of generalization from the learned set of patterns.
{
As it will be shown in the next section,
 the phase precession
as the animal moves from one location to the other is related to this
interpolation mechanism. 
}

An example of learning kernel $A(\tau)$ that satisfy the
resonance conditions (\ref{Constraints})
at frequency $\omega_0=10$ Hz (theta range) is shown in Fig. 1.
This learning kernel is able
to imprint oscillations with frequency $\omega_0=10$ Hz,
and to make the network resonant at theta frequencies.
It could be of interest to note that 
the kernel $A(\tau)$ shown in Fig. 1 
satisfies the resonance condition also at $\omega_0=40 Hz$, i.e. in the gamma range.
However we will not exploit this last property in this paper.

}

   \section{ 
 Imprinting locations
and the theta phase precession } 
 Let's consider N locations, $x_\mu$ arranged in a 1-dimensional 
space with periodic boundary
      conditions 
{(such as in a circular or triangular track). 
For each of these N  locations $x_\mu$, $\mu=1,..,N$
we imprint, in the synaptic strengths of the network,
 an oscillating  pattern $\bP^\mu$, with specific amplitude
 and phase relationship between its components. 
Using eq. (\ref{rule}) in order to imprint N (linearly independent) patterns,
 one need a number of network units $n>N$.
In the following we use a network with n=2N excitatory units
 and the same number of
 inhibitory units,
that follows the equation of motion (\ref{eqnsv}). 
}
 {
For each location $x_\mu$ an oscillating pattern
$P^\mu$, whose amplitude profile is peaked in the excitatory unit ${2\mu}$,
is imprinted in the network.
}
 {
   In particular, the imprinted pattern
 associated with location $x_\mu$ has components
}
$P_j^\mu(t) = \xi_j^\mu \e^{-\i \omega_0 t} + \cc$,
 $\, \mu=1,2,..,N$, $j=1,2,..,2N$,
  where $\xi_j^\mu = \A(2\mu -j) e^{i\phi (2\mu - j)}$;
the  {amplitudes} ${\A}$ and  {the phases}
 $\phi$
 of the pattern $\bxi^i\mu$ on each unit $j$
 are functions of the distance
 {
between the unit $j$ and the 
unit  $2\mu$ that is associated with location $x_\mu$.
} 
 In particular, $\A( )$ is an even function
   decreasing quickly with distance, while $\phi ( )$ is an odd function.
   This means, when imprinting a location $x_\mu$, only units
   located close to unit $2\mu$ are excited by the imprinting
   inputs.
 {
In such a manner, after the learning process,
 the N excitatory units $u_{2\mu}$, that we will call place cells,
 will be associated with a receptive field
centered at $x_\mu$, for $\mu=1,2,\ldots,N$. 
The other N units (with odd index, called auxiliary units)
are
 necessary in order to allow the network to store the
N desired patterns properly.}
   In all simulations  
we used 
 {
$\A(0)=1$, $\A(\pm 1)=0.3 $ $\A(\pm 2) =0.3 $, $\A(x)=0$ otherwise,
   and $\phi(0)=0$, $\phi( \pm 1)= {\mp } 2.4 \,rad$ and $\phi(\pm 2)= \mp 2.5\, rad$.
   The network has a quite robust behavior with respect to the choice 
   of the functions $\A$ and $\phi$.
}

    {
 In our model the external inputs in the recall mode
   resembles but are
    not exactly the imprinted patterns. }
When the animal is exactly at location $x_\mu$, 
the input pattern $\delta \bI$ is given by
    $\delta \bI = \L^\mu \e^{-\i \omega_0 t}$ where 
$\L^\mu$ is a vector with components $L_j^\mu=\delta_{j,2\mu}$.
    This pattern conveys the animal location in the amplitude modulation but has no
    phase  {coded} information. 
    When the animal is at position x between  location 
  $x_\mu$ and %
location $x_{\mu+1}$,
    $\, x_\mu < x  < x_{\mu+1}$, 
 we take the input vector as linear combination of $\L^\mu$ and $\L^{\mu+1}$,
\be
\L^x= (1-|x-x_\mu|) \L^\mu + |x-x_\mu| \L^{\mu+1}
\label{lx}
\ee
  { L is then normalized. 
Thus, if the animal moves from location $x_\mu$ at time $t=0$ to the adjacent 
one $x_{\mu +1}$ with speed $v(t)$, the current location
will be a function of time $x(t)= x_\mu+\int_0^t v(t) dt$, 
and      the input  vector will be expressed by eq. (\ref{lx})
where x is replaced by $x(t)$. 
The vector $\L$ is always normalized.    
 Fig. \ref{fdynamics}B shows  
the input amplitude $L_j(t)$ when animal moves along the track with constant speed.
     }

{
    A animal at location $x$ receives an input pattern $L^x$
    that resembles imprinted pattern $\bxi^\mu$ for $x_\mu \approx x$.
    Hence, the network response to input $\L^x$ would amplify
    the imprinted  {pattern} $\bxi^\mu $ that
    resembles $\L^x$ most (see previous section), and thus
    exhibits in the responses  the non-trivial phase relationship among units
   that was stored in the couplings.
  When animal is in between two stored locations, 
  the response will be a combination of the
two imprinted patterns (see eq. (\ref{ii})), and
in particular, the phase of 
the response will be a combination
of the phases of the two imprinted overlapping patterns
which depends from the 
animal position.
     So, even though the input carries information about current 
    location only in the  amplitudes modulation,
    the associative memory nature of our model,
having phase-coded locations imprinted in the synaptic couplings, 
    allows  the network to 
    show both amplitude and phase coded  informations in the response.
}

   Since the theta rhythm is the local field potential,
 in our model it is simply the mean field.
We  compute the theta rhythm doing the average of all the
  excitatory units activities $ M.F.= 1/2N \sum_{j=1}^{2N} u_j(t)$.
    This choice allows us to compare the the absolute phase of theta at which
place cells fire during the place field traversal, with the
experimental findings of Skaggs et al. (1996).
Indeed,
 the phase zero of the theta rhythm in the experimental work of Skaggs et al 1996
 is
defined to be the point in the theta cycle corresponding to the
maximal pyramidal cells population activity (i.e.
the average over the entire
dataset of pyramidal cells), and 
the EEG is opportunely shifted  in order to agree with this definition of
theta rhythm phase.

    \section{Simulation Results and comparisons with experimental findings}
    In all simulations reported here we used the kernel $A(t)$
    shown in fig. \ref{kernel} that satisfy the
     constraints (\ref{Constraints}) when  {$\omega_0= 10 Hz$,
    i.e. when the frequency is in the theta range.}

After having imprinted $N=10$ locations  in the net with $n=20$ excitatory units,
 we simulate the network dynamics while the animal moves
continuously in space.
Fig. \ref{fdynamics} shows in red the activity of
the two place cell units ($u_{2\mu}$ with $\mu=5,6$),
 as a function of time,
while the animal moves with constant running speed.
Black curves shows the theta rhythm computed
as the mean field of all the excitatory units of the network.
 The simulation is done at a constant animal speed 
$v=1/ (400ms)$
(note that space is adimensional in our framework, 
and a unitary distance correspond to the distance
between the centers of two adjacent receptive fields).
The oscillatory input on each unit $j$ is $I_j(t)= L_j(t) cos(\omega_0 t)$,
where the amplitude $L_j(t)$ is shown in fig. \ref{fdynamics}B
as a function of time.
The figure shows that, for
 each place cell, the oscillation amplitude of the firing rate
increases as the animal approaches the center
of its place field and decrease as the animal left the center
 (firing tuning curve),
while the place cell phase advanced at consistently 
earlier phases of the theta cycle
as the animal pass through the cell's receptive field (phase precession). 
To quantify the phase precession phenomena, 
we compute the phase shift between the theta rhythm and
the place cell activity in each cycle, and we plot in fig. \ref{fdynamics}
the theta phase  as a function of the animal position 
during running, together with the amplitude modulation.

The model dynamics is in agreement  with the  experimental evidence of theta phase
precession  in two main points:
1) the total phase precession is  {less than} $360^o$ 
(O'Keefe Recce 1993, 
Skaggs 1996);
2) as the animal enters the receptive field the
absolute phase of the cell activity with respect to
the phase of maximal excitatory units population activity (theta rhythm) 
is slightly less than $180^o$,  in agreement 
with experimental finding of Skaggs et al (1996).
Effects on nonlinearity in our general model have been
analyzed in  Scarpetta et al (2002), where we distinguished two class
of nonlinearity.
For the nonlinearity of class I (as the sigmoidal function
shown in fig. \ref{fdynamicsNONLIN}B) we expect
 that the nonlinearity do not change critically the linear results.
Indeed comparing the linear case in fig.  \ref{fdynamics}
with the nonlinear simulations results shown
 fig. \ref{fdynamicsNONLIN} we see that
the dynamics is affected by the nonlinearity  at large
amplitudes but the
phase precession is very well preserved.

To check if in our model the theta phase is better
correlated with position than with time,
we simulate the network dynamics with different values
of the animal running speed.
In fig. \ref{ff}  the theta phase is shown as a function of
animal position when animal runs with three different 
speeds ($v=1/800 ms^{-1}$ black squares, 
 $v=1/400 ms^{-1}$ red triangles, 
 and $v=1/200 ms^{-1}$ green circles).
As we expect,
the theta phase is a unique function of the animal position,
almost insensitive to the running speed,
 as long as the reaction time 
 of the network is short compared with
the time characteristic of the animal movement.
 While, when the speed is  large
the inertia-like effects  becomes significant
and a (small) dependence from the speed appears
(green circles follow a line with a higher slope)
 in the second half of the receptive field.
From figure \ref{ff} we see that, when there's variability in running speed,
the phase shows a better correlation with position than with time.
This is in agreement with experimental results
os O'Keefe Recce (1993) and Huxter et al (2003).

{
Finally, in order to test our model,
we compare it with the
recent experiments of Zugaro et al 2005.
We have simulated the
effect of a perturbation 
that silences all units
(both excitatory and inhibitory ones)
and  reset the phase of theta rhythm.
The activity of two place cells, before during and after
the network perturbation, is shown in fig. \ref{ff3}.
The perturbation last for 200 ms.
Despite the theta phase reset and the transient interruption of
firing, 
the theta phase of 
 the first  cycle after the recovery is
 more advanced than the theta phase of the last cycle
before the perturbation,
as in the experiments 
(Zugaro et al 2005).
The decrease of  the peak firing rate is stronger in our simulations than
in the experimental findings of Zugaro et al (2005),
 that reported only
a small but not-significant decrease in peak firing rate.
To quantify the
degree of conservation of the phase precession 
and the effect on the amplitudes of oscillation
we shows in fig. \ref{ff3} the plot of the
theta phase as a function of the position, and
the amplitude of the oscillation versus the animal position.
Figure \ref{ff3}C should be compared with fig. 2 in Zugaro et al (2005).
}

 {
Figs. \ref{fj} and \ref{fw} show the synaptic connections $J_{ik}$
and $W_{ij}$ learned in our model,
as a function of pre-synaptic cell index k, for a post-synaptic place cells (i even)
and an auxiliary units (i odd).
As expected, the synaptic connections are translation invariant,
so $J_{ij}$ depend only on the difference $i-j$ and 
on the parity of $i$ (even or odd).
We note that connections  have both positive
and negative values. Since they are supposed to be 
projections from excitatory pyramidal cells, the negative connections weights means
effectively
inhibitory connections, which could be implemented 
by additional inhibitory interneurons
with very short membrane time constant.
Note that the synaptic weights are highly asymmetric.

Finally, note that in our model the activity of all the units, both
 excitatory and inhibitory units, is modulated at the theta frequency.
The network shows resonance in the theta range, and
the inhibitory units, similarly to the place cells,
 exhibit  phase precession (not shown),
while the auxiliary units have a very weak place field with very little
phase precession (not shown).

\section{Conclusions and Discussion}

 {
In this paper we propose a  theoretical model
for theta phase precession in the hippocampus
by applying a framework (Scarpetta Zhaoping Hertz 2002)
that is a generalization of the Hopfield model
to %
 oscillatory patterns. 
The learning rule is a generalization of the Hebb prescription,
inspired to the STDP synaptic plasticity.

 The associative memory behaviour of our model,
    allows  the network to
    show both amplitude and phase coded informations in response
to inputs that carry information about current
    location only in the  amplitudes modulation.
}

Our model dynamics is in agreement  with the  experimental 
evidence of theta phase
precession  in 3 main points:{

1) the model accounts for the observations that 
after a perturbation that transiently turned off neuronal discharges for
a couple of theta cycles and reset the phase of the theta oscillations,
 the phase precession continued (Zugaro et al.  2005).
This happens  in our model, even if the external input
to the network is silenced during the perturbation.
This comes naturally 
from the fact that in our model,
the oscillation phase of a place cell depends
on both the network connections and the external input strength.
Hence, when the external input strength represents the spatial
location (relative to the center of the place field),
the oscillation phase of the place cell will, by the influence of the
network connections, code the
location relative to the center of the place field.

2) 
In our model 
the theta phase depends on current position
of the animal, and is 
only slightly sensitive to the
animal speed  (almost insensitive at low speeds).
 Therefore in our model 
the theta phase is  better correlated with position
than with time spent in the place field,
 in agreement with experimental data 
(O'Keefe Recce 1993, Huxter et al 2003).
This is achieved
without need of any speed-dependent input,
in contrast with the recent model of phase precession 
(Lengyel et al  2003)
that  assumes a differential input proportional to
the instantaneous velocity of the rat. 

3) Both the  initial theta phase as the animal enters the place field
and the total phase precession generated by our network dynamics,
 are in quantitative agreement 
with experimental evidence.
In our model the total phase precession is less than $360^o$,
and the absolute phase of the cell activity w.r.t. the
average pyramidal cells activity, when the animal 
enters the place field, is slightly less than $180^o$,
in agreement with experiments of Skaggs et al. (1996).
Note that 
our
definition of the theta rhythm as the average 
activity of all the excitatory units
is in agreement with
 the definition used   by Skaggs et al (1996). 
}

Finally note also that our model predicts
both a monotonic phase  precession
and a unimodal firing tuning curve with a large amplification
factor when the animal is in the center of the place field.
This is in contrast with some other models.
For example,
models in which both firing rate and phase
are linear or monotonic functions of the same parameter
(Booth and Bose 2001; Magee 2001) produce
a monotonically increasing firing rate inside the place field
if monotonic phase precession is preserved,
or non-monotonic phase precession if unimodality
of the firing profile is to be preserved.

Although 
 there is some controversy as to whether
theta phase precession is linear (O'Keefe Recce 1993)
 or if
theta phase is an accelerating 
function of position (Skaggs et al 1996, Yamagushi et al 2002),
and if it's smooth or clusterized,  its
trend is undoubtedly monotonic (O'Keefe and Recce 1993; Skaggs et al 1996,
Yamagushi 2002), since there is a systematic progressive phase
retardation as the animal passes through the place field. 
 
In contrast, 
 experimental data indicates that firing rate
 changed in a non-monotonic
waxing-waning manner (O'Keefe Recce 1993, Skaggs et al. 1996),
even though there is some controversy as to 
whether phase precession accompanies skewed firing
profiles (Mehta et al.
2002; Harris et al. 2002) or not (Huxter et al. 2003).

Our model shares some similarities to the one of Tsodyks et al 1996,
since both are based on the associative memory properties of the
network. In contrast with the Tsodyks et al (1996) work, here
the couplings are computed using a
mathematical learning rule
and the network dynamics is studied
analytically in a framework  (Scarpetta  et al 2002)
that is the generalization
of the Hopfield model to the case of  not-static 
but oscillatory patterns.
A point of our model that still needs  investigation is the
origin of the imprinted patterns used during the
learning mode.
Of course, if
 there is no phase differences between units in the imprinted patterns,
there would be no phase precession despite of learning and network dynamics.
We are currently investigating how such  phase
differences arise in the imprinted patterns during learning,
i.e. 
 if they are generated by an initial asymmetry of
the connections that produces the phase shift among units when the 
animal is moving,
or if they are inherited from earlier stages input during learning.
Anyway, what we have proven here is that when
 the network
is equipped with such a synaptic connectivity
the network dynamics 
shows phase precession and phase-coded response to sensory input
that do not have any phase difference between units activities.
}

It has  previously been suggested that phase coding
occurs in the sensory input to
the place cells (Burgess et al. 1994), and it has been shown
that under that model individual place cells
will show phase precession versus the average
place cell activity (Burgess et al. 1993), as in 
our model. 
A difference with their model is that,
even if we use a phase-coded activity during learning,
once the network is equipped with the proper synaptic couplings,
the phase precession occurs, when the animal run, without need
of any phase coding in the sensory input.

\section*{Acknowledgements}
The idea of this work originated from a discussion with L. Zhaoping.
Authors would like to thank her for the useful discussions, comments,
and the critical reading of the manuscript.
We would like to thank also the anonymous referees for
the useful criticisms and 
comments which allow us to improve the paper.
This work was partially supported by INFN project FB11.

\begin{figure}[h]
\begin{center}
\setlength{\unitlength}{1cm}
\begin{picture}(9,3)
\put(0,0){\includegraphics[width=3cm]{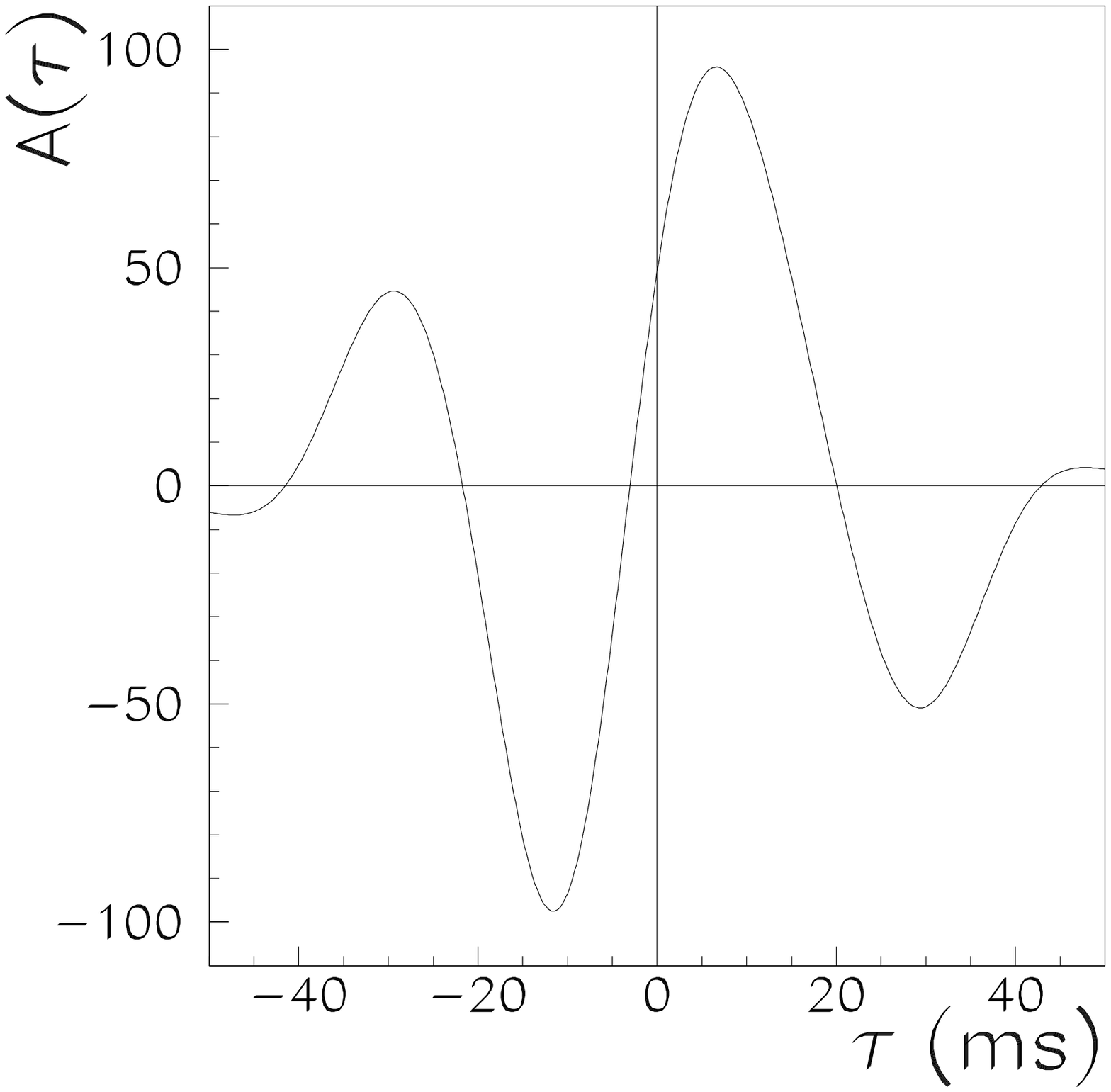}}
\put(0,0){A}
\put(3,0){\includegraphics[width=3cm]{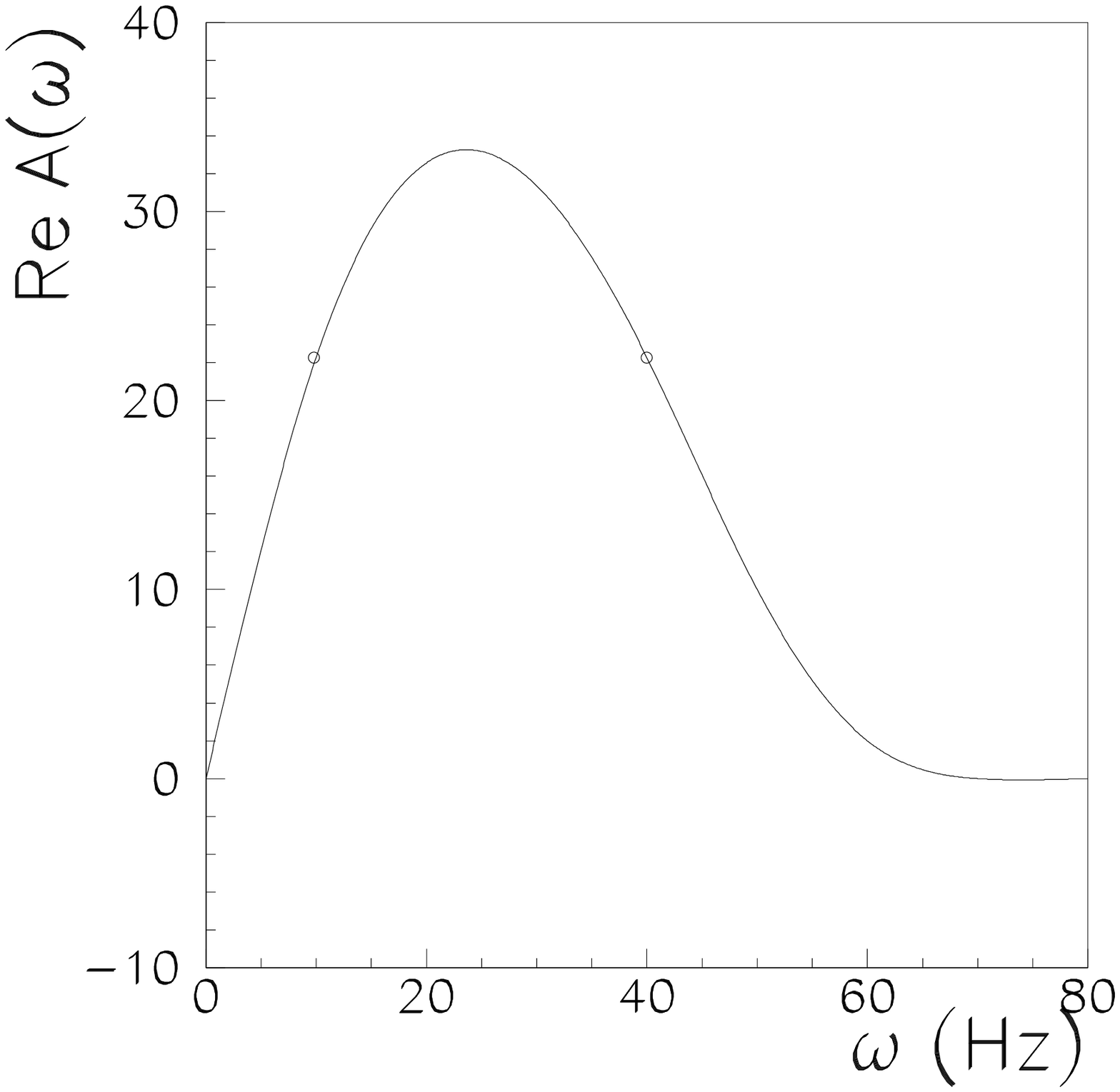}}
\put(3,0){B}
\put(6,0){\includegraphics[width=3cm]{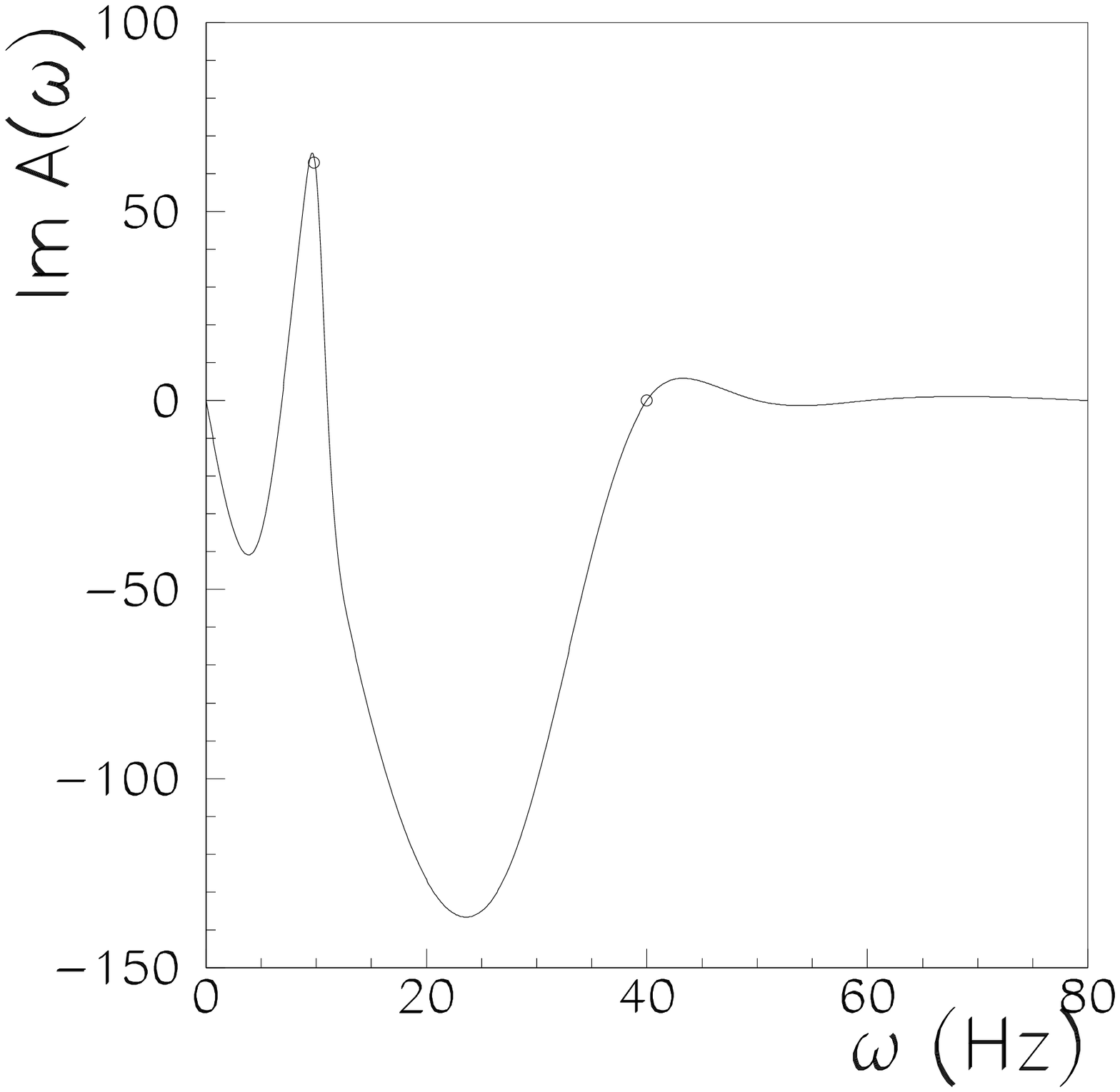}}
\put(6,0){C}
\end{picture}
\end{center}
\caption{A. Example of a  learning kernel $A(\tau)$ that satisfy the resonance
conditions \ref{constraints},  at   frequency $\omega_0=10 Hz$.
(Parameters $\gamma=\beta=0.2 ms^{-1}$, $\alpha=0.14 ms^{-1}$).
The real (B) and the imaginary (C) part of its Fourier transform. 
In all the
simulations we used the learning kernel shown in A.
}
\label{kernel}
\end{figure}

\begin{figure}[h]
\begin{center}
\setlength{\unitlength}{1cm}
\begin{picture}(9,6)
\put(0,3){\includegraphics[width=9cm]{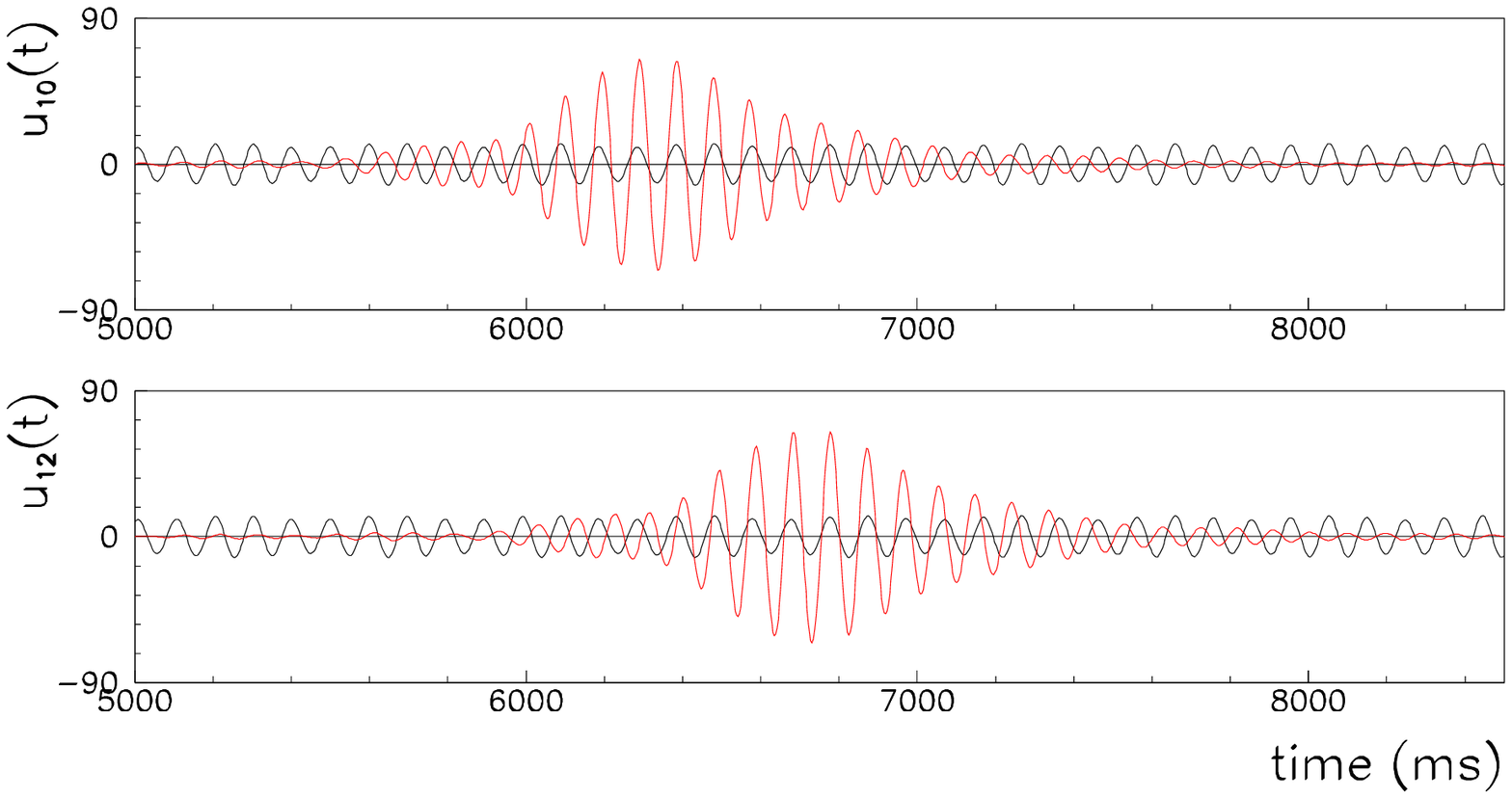}}
\put(0,3){A}
\put(0,0){\includegraphics[width=4.2cm]{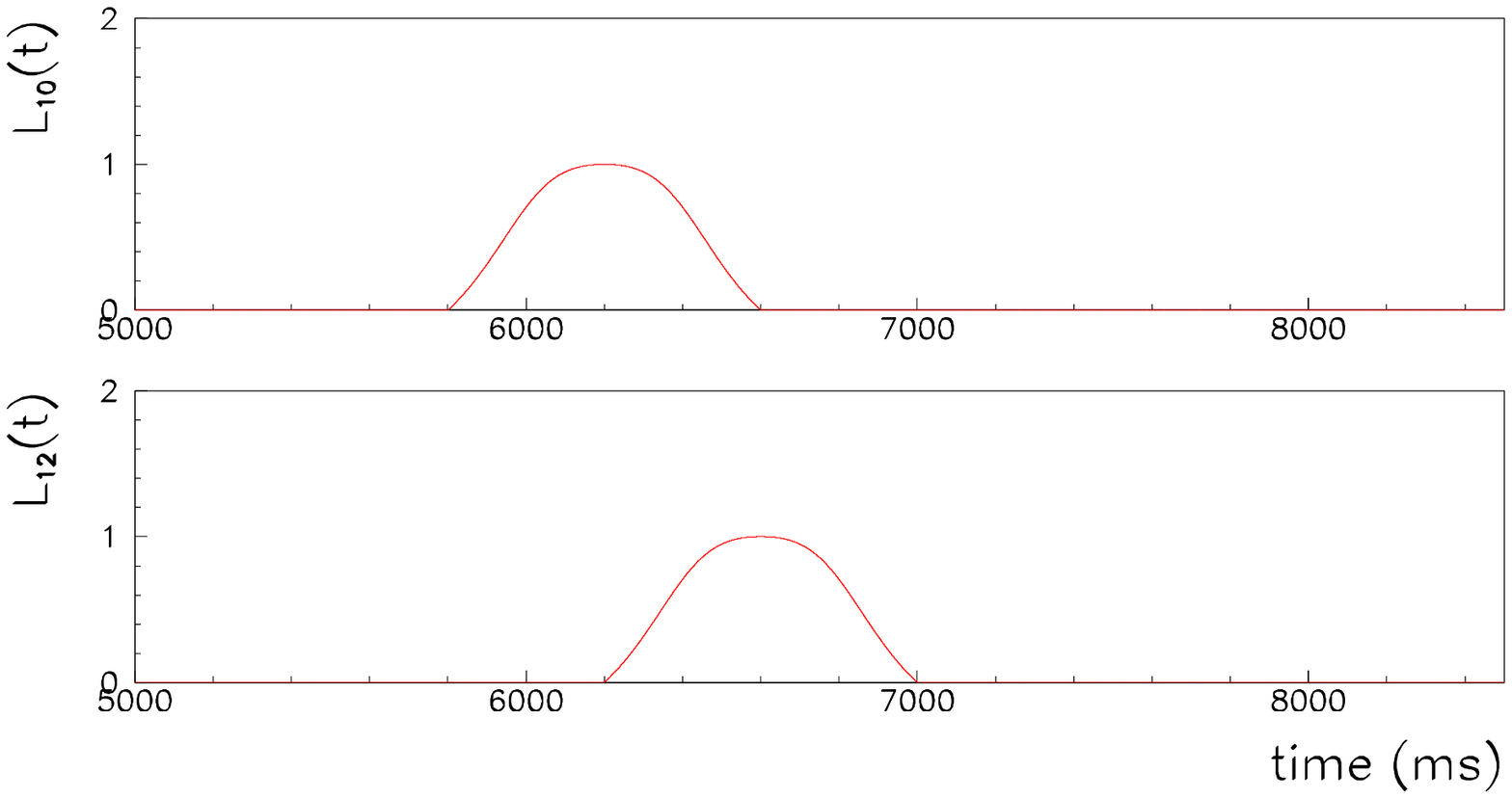}}
\put(0,0){B}
\put(4.5,0){\includegraphics[width=4.5cm]{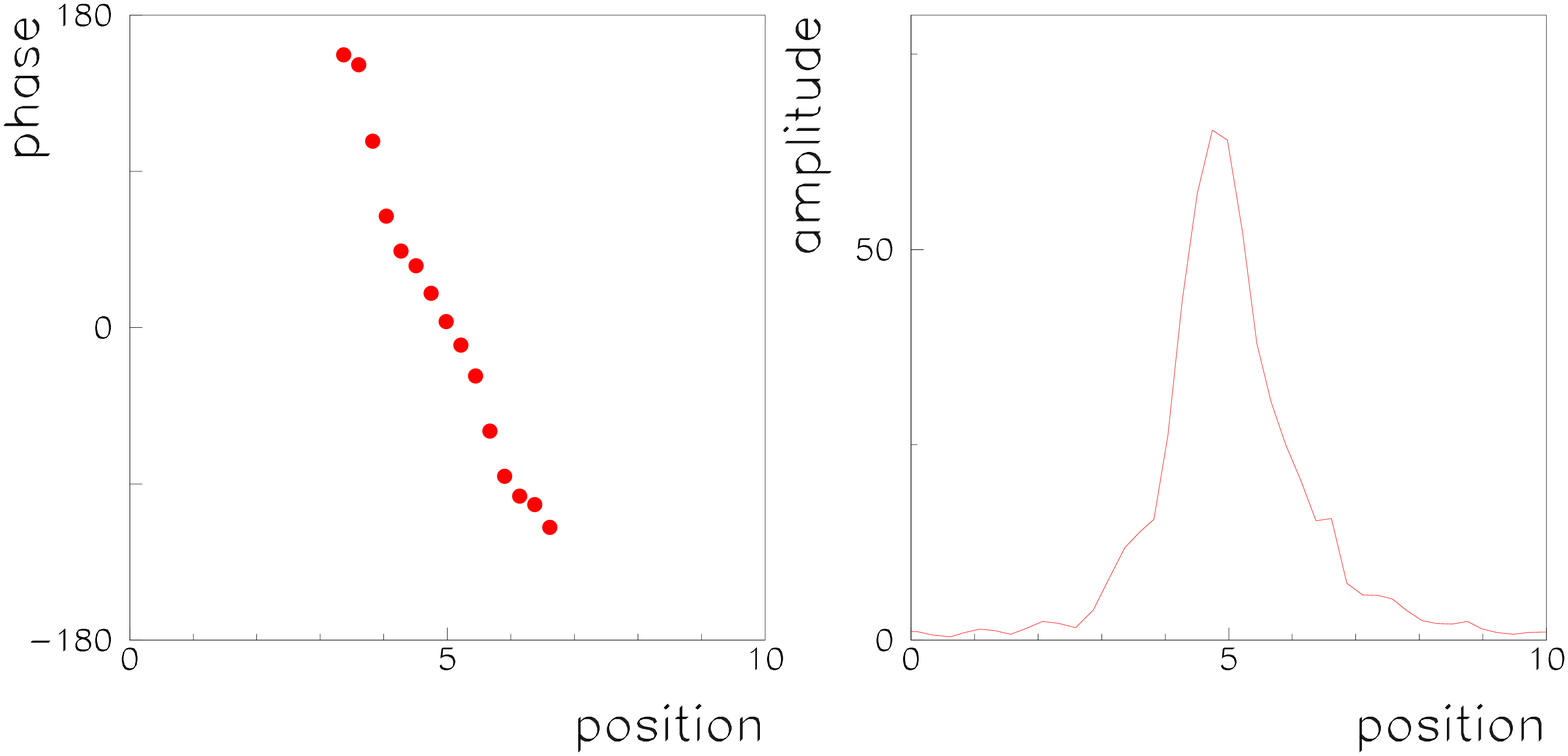}}
\put(4.5,0){C}
\end{picture}
\end{center}
\caption{%
 Simulation of the network dynamics when
animal is moving continuously in time with constant 
running speed v=1/(200 ms).  
A: The two subplots show the activities (in dotted red) of two place cell units
($u_j$, $j=10,12$, whose
place fields are centered at $5$ and $6$ respectively) as a function of time.
Black curves show  the ongoing theta rhythm activity
(computed as the mean field
of the excitatory activity of all the network).
 The 
external input to the units is $I_j(t)=L_j(t) cos(\omega_0 t)$,
where the amplitude $L_j(t)$, shown in B, 
is maximum when the animal is in the center
of the receptive field of that unit j. 
Note that each place cell unit shows phase precession, i.e., relative
phase shift between $u(t)$ (red) and the theta activity (black), as
the animal enters and then leaves the place field of this unit.
Same behaviour for all other place cells not shown.
C: Theta phase as a function of the current position of
the animal,  and amplitude of oscillation
as a function of current animal position,
 for the place cell centered in location $5$.
Theta phase is a bit less than $180^o$ when the
animal enters the receptive field, and the
total phase precession is  less than $360^o$.
}
\label{fdynamics}
\end{figure}


\begin{figure}[h]
\begin{center}
A\includegraphics[width=7cm]{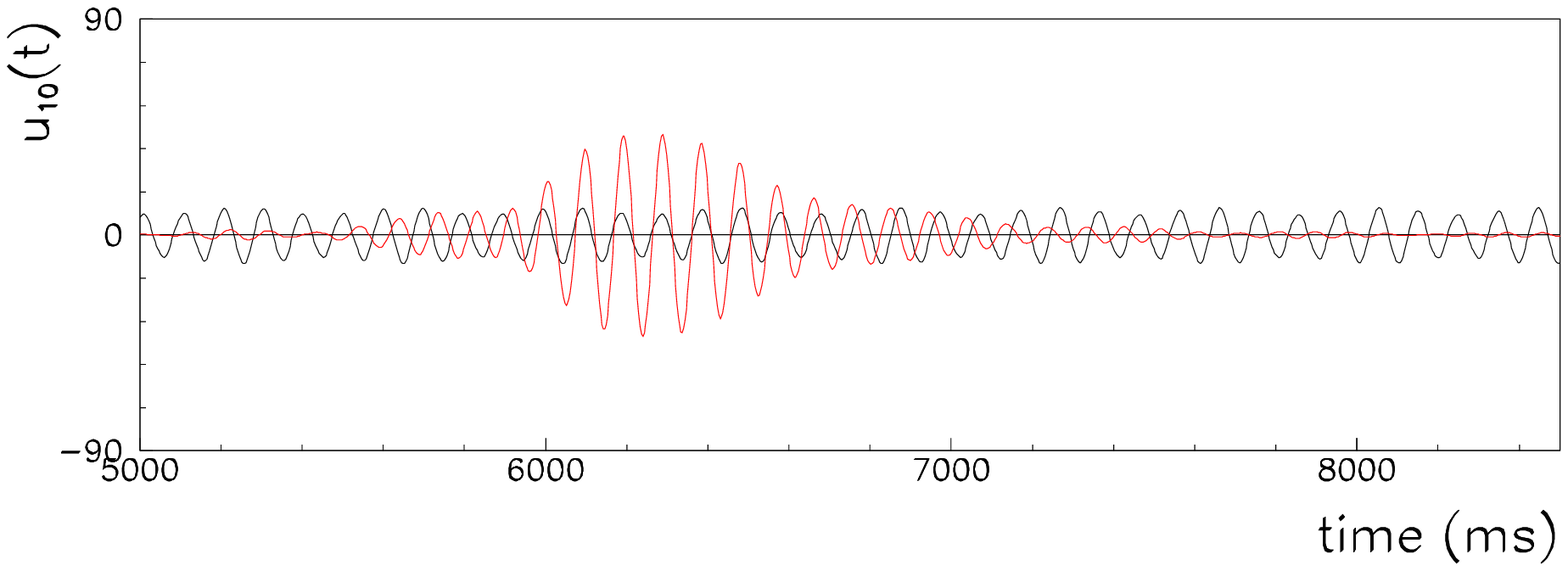}
B\includegraphics[width=2cm]{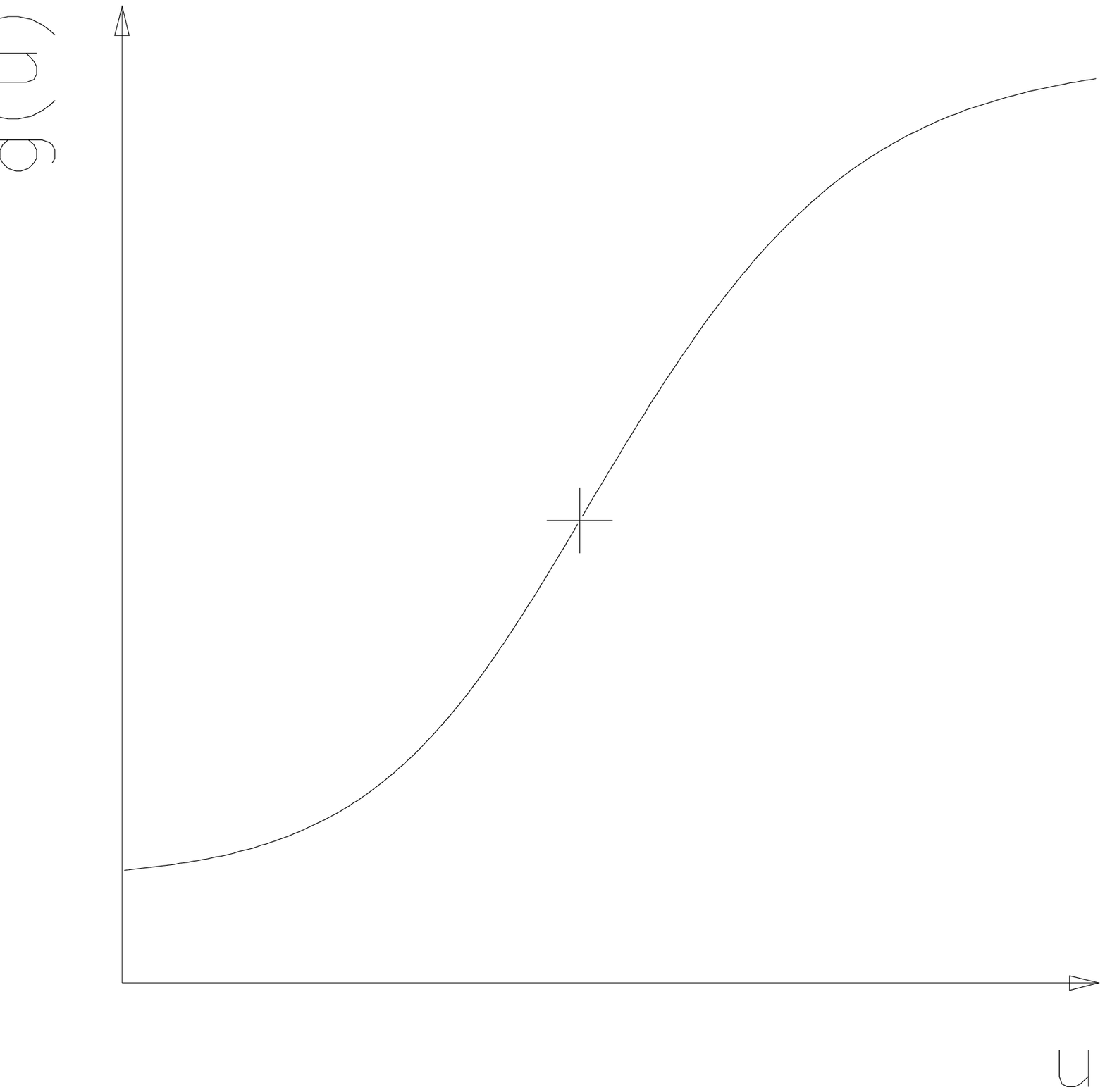}
C\includegraphics[width=5cm]{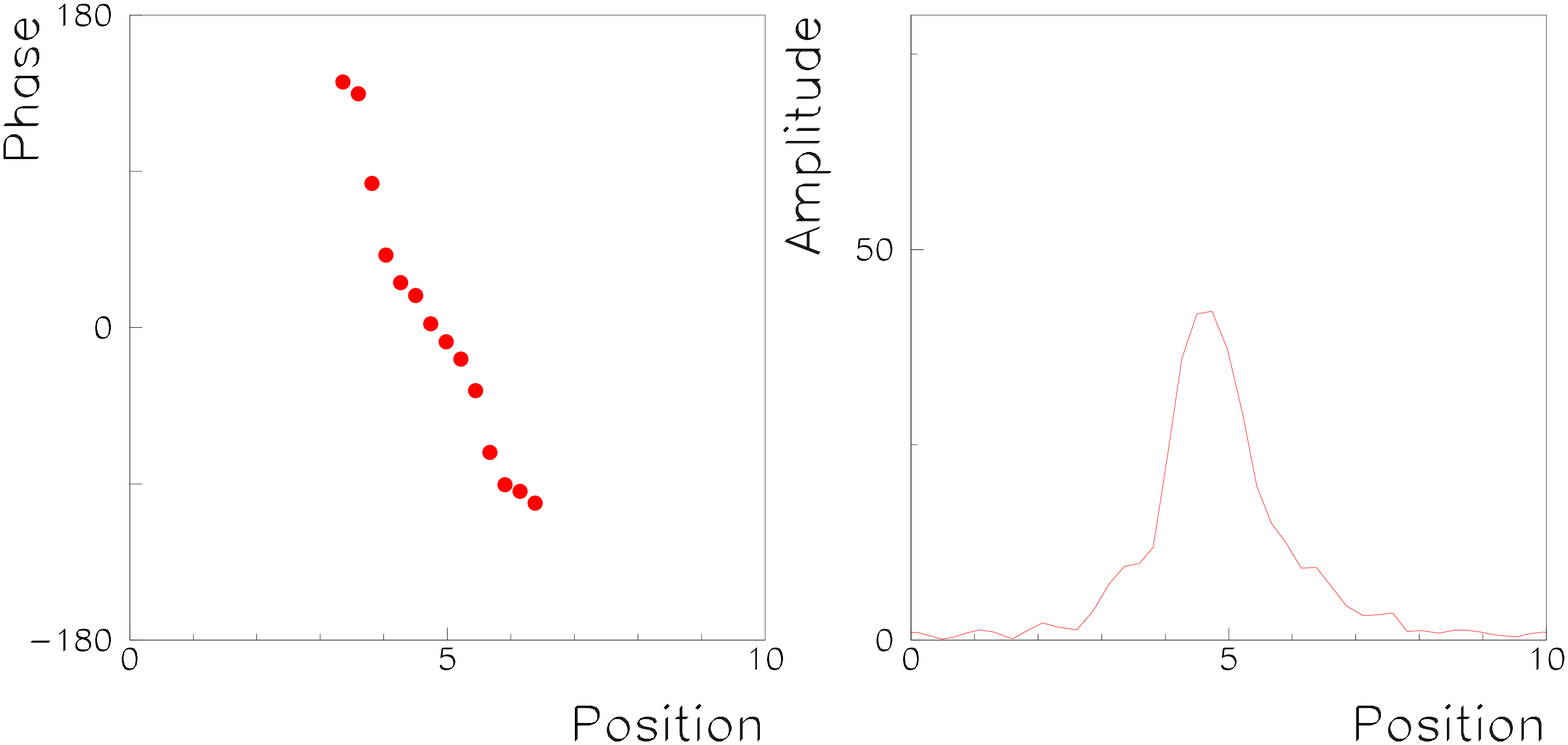}
\end{center}
\caption{%
Effects of nonlinearity in the phase precession phenomena dynamics.
 A. Simulation of the network dynamics using the nonlinear activation
function showed in B, when
animal is moving continuously in time with constant 
velocity v=1/(200 ms) as in previous figure.
Dotted red curve shows the place cell activity 
as a function of time.
Black curves show  the theta rhythm.
Results are affected by nonlinearity mainly at
large amplitude, but the theta phase precession is preserved, as shown in figure
C showing the theta phase as a function of position, and amplitude of oscillation
as a function of position for the place cell with receptive field centered in 5.
 }
\label{fdynamicsNONLIN}
\end{figure}

\begin{figure}[h]
\begin{center}
\setlength{\unitlength}{1cm}
\begin{picture}(9,6)
\put(0,0){\includegraphics[width=4cm]{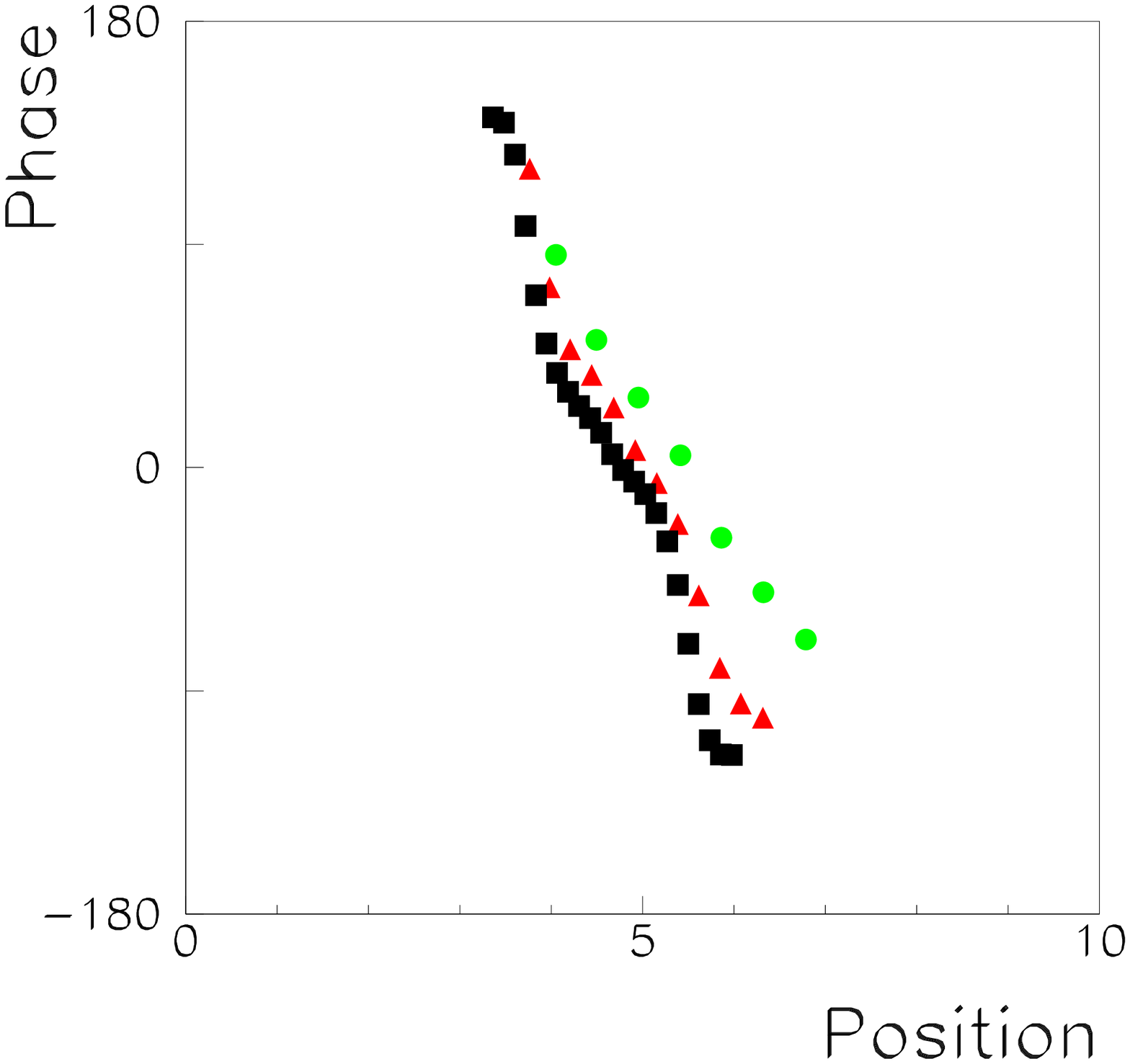}}
\put(0,0){A}
\put(4.5,0){\includegraphics[width=4cm]{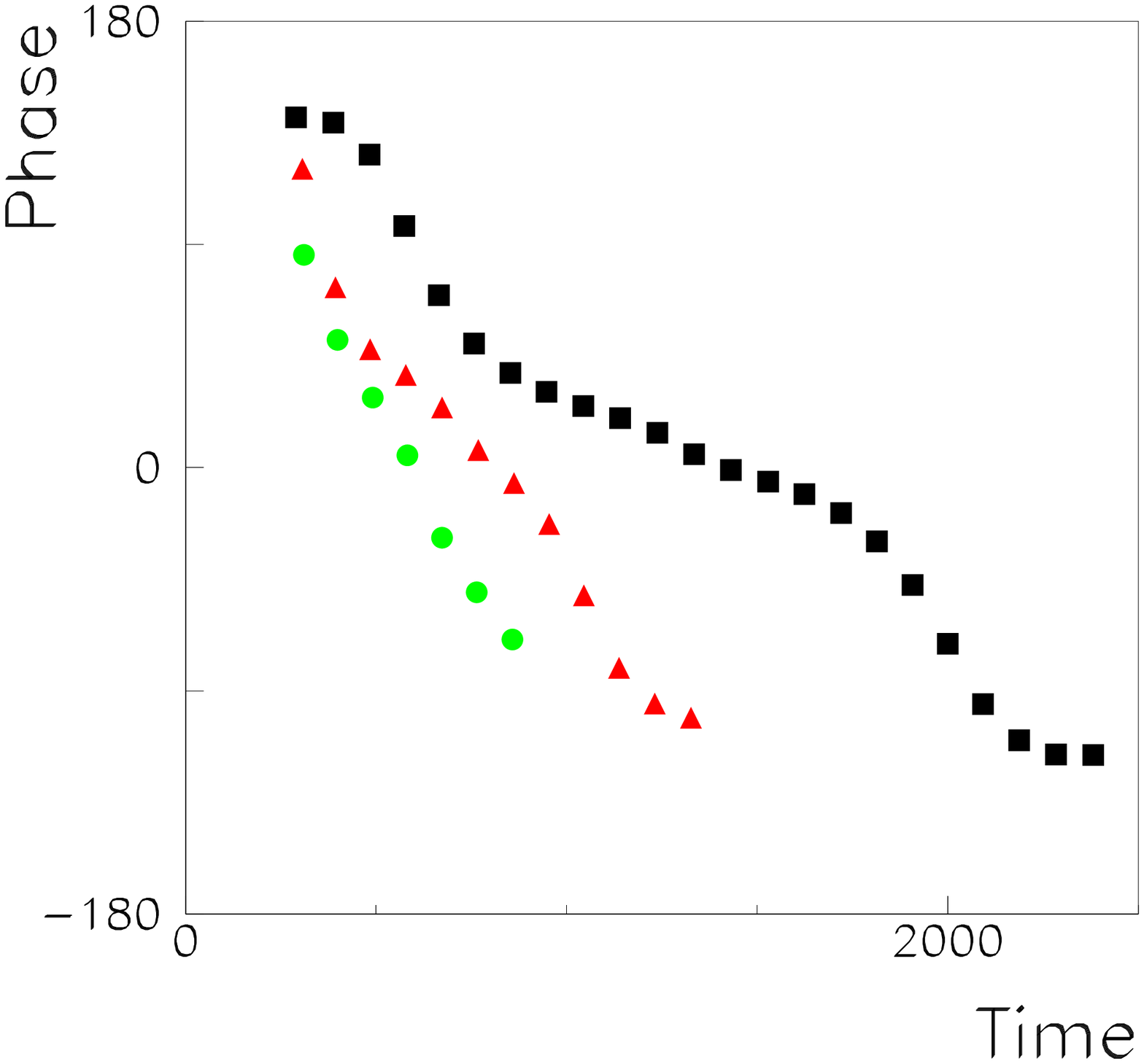}}
\put(4.5,0){B}
\end{picture}
\end{center}
\caption{%
Theta phase of the place cell as a function of the animal position (A),
and as a function of the time spent in the place field (B)
at three different animal running speeds.
During the place field traversal, for each cycle,
 we compute the phase of the place cell  
with respect to the theta rhythm.
In each of the three simulations, the
 animal is moving continuously in time with constant 
speed. The green circles correspond to speed v=1/200 $ms^{-1}$,
the red tringles to v=1/400 $ms^{-1}$, and the black squares to v=1/800 $ms^{-1}$. 
Figure shows that the theta phase is much better correlated with
position than with time.
}
\label{ff}
\end{figure}

\begin{figure*}[h]
\begin{center}
\setlength{\unitlength}{1cm}
\begin{picture}(16,9)
\put(4,7.2){\includegraphics[width=8cm]{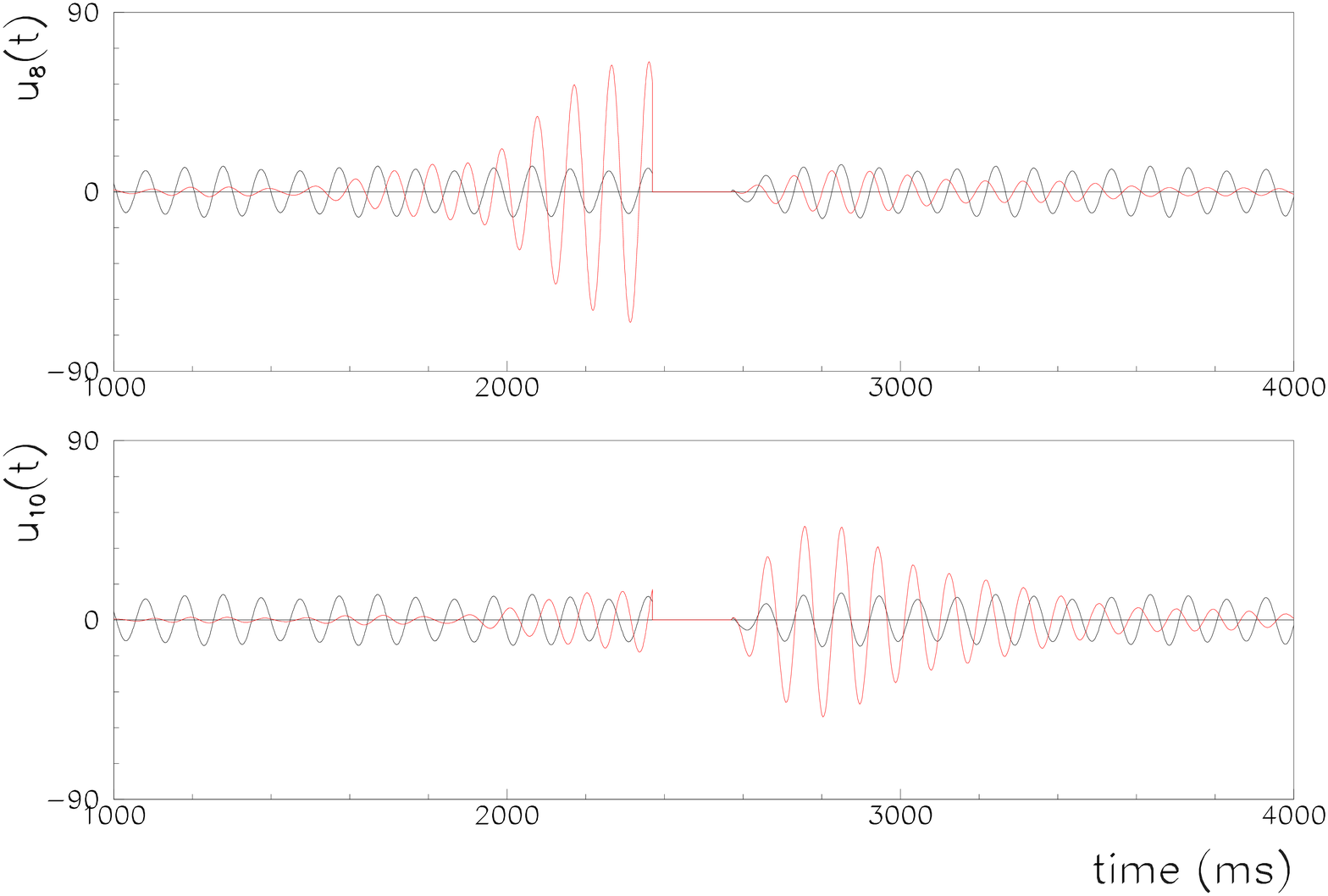}}
\put(4,7.2){A}
\put(3,6.4){With perturbation}
\put(1,3.4){\includegraphics[width=6cm]{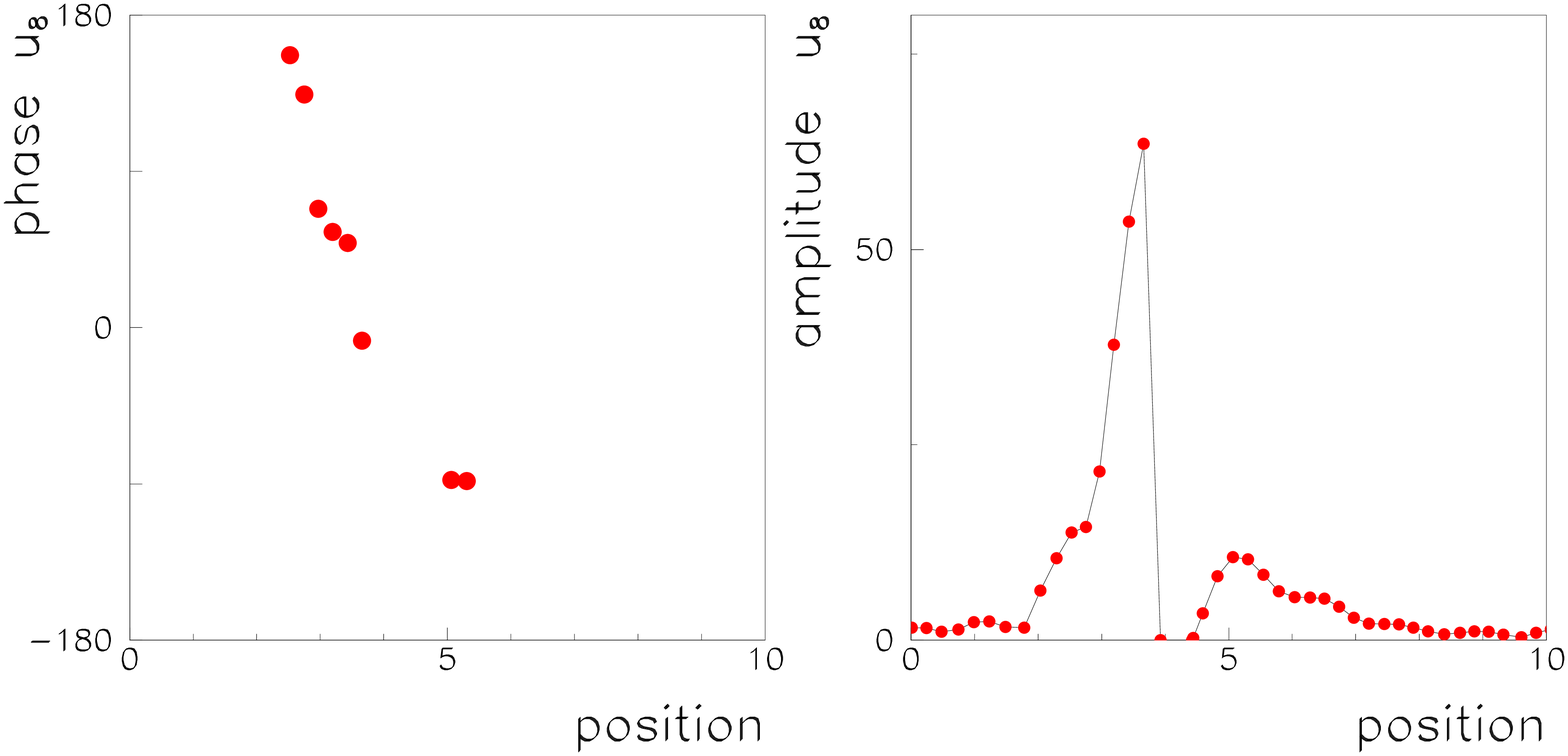}}
\put(3.5,3){Control}
\put(1,0){\includegraphics[width=6cm]{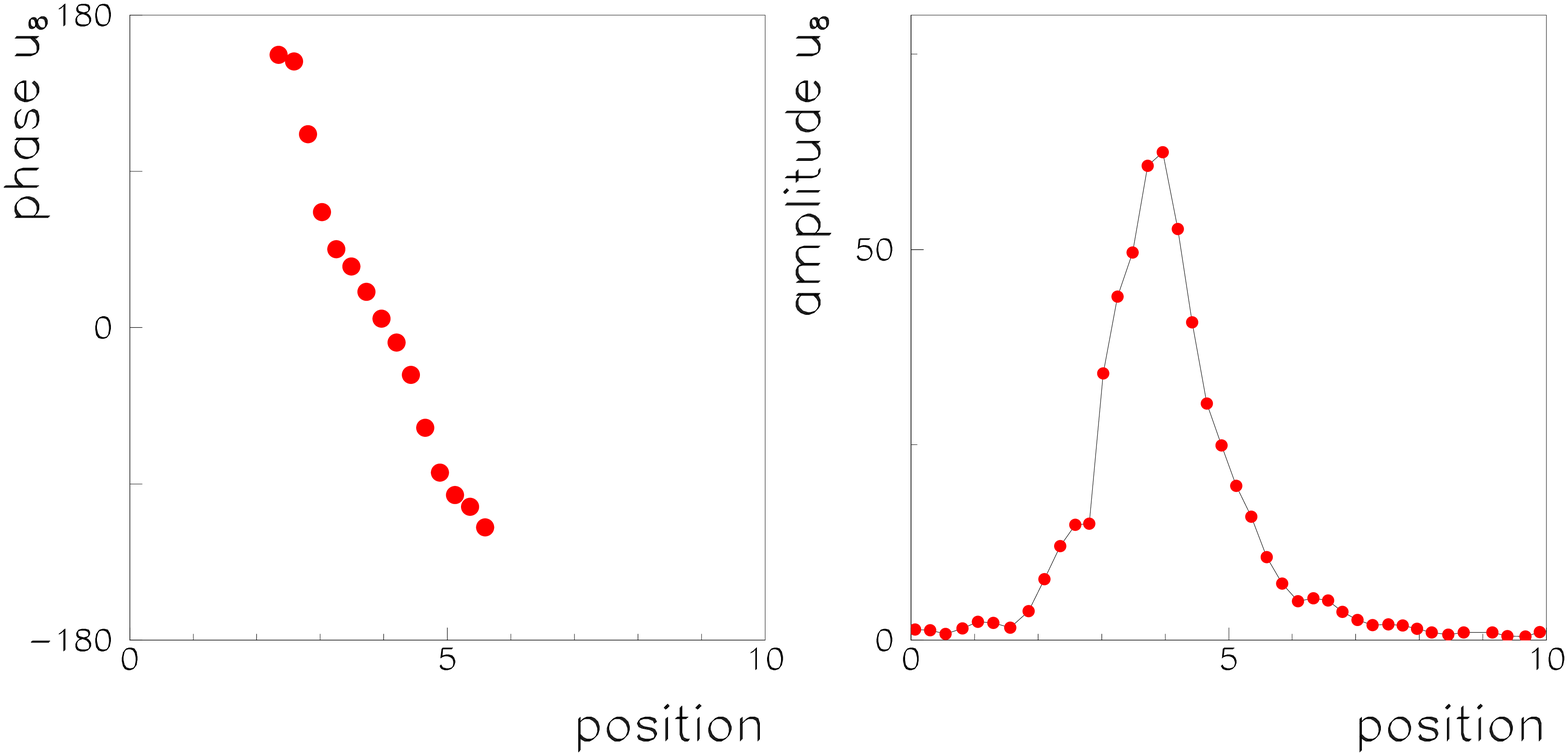}}
\put(0,0){B}
\put(11,6.4){With perturbation}
\put(9,3.4){\includegraphics[width=6cm]{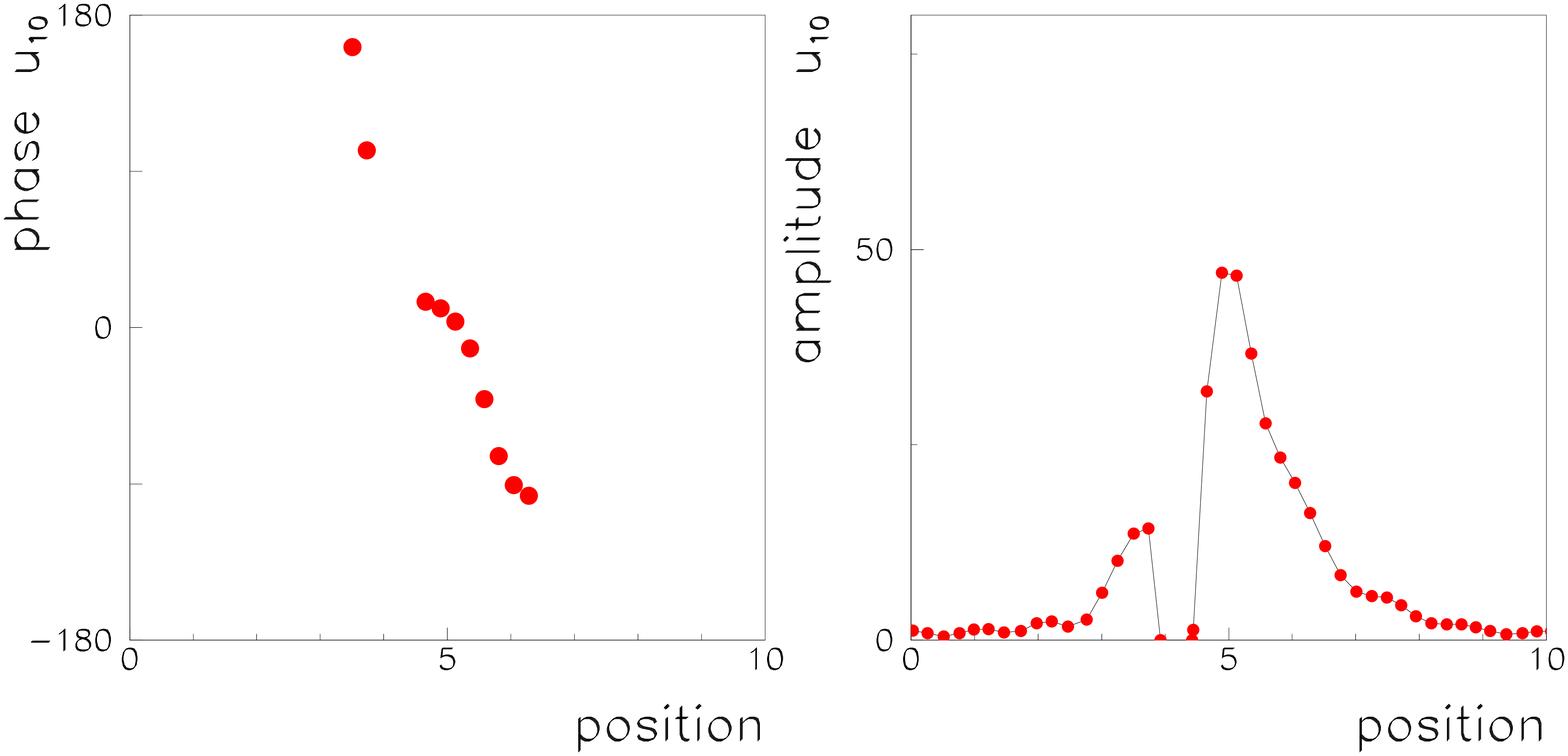}}
\put(11.5,3){Control}
\put(9,0){\includegraphics[width=6cm]{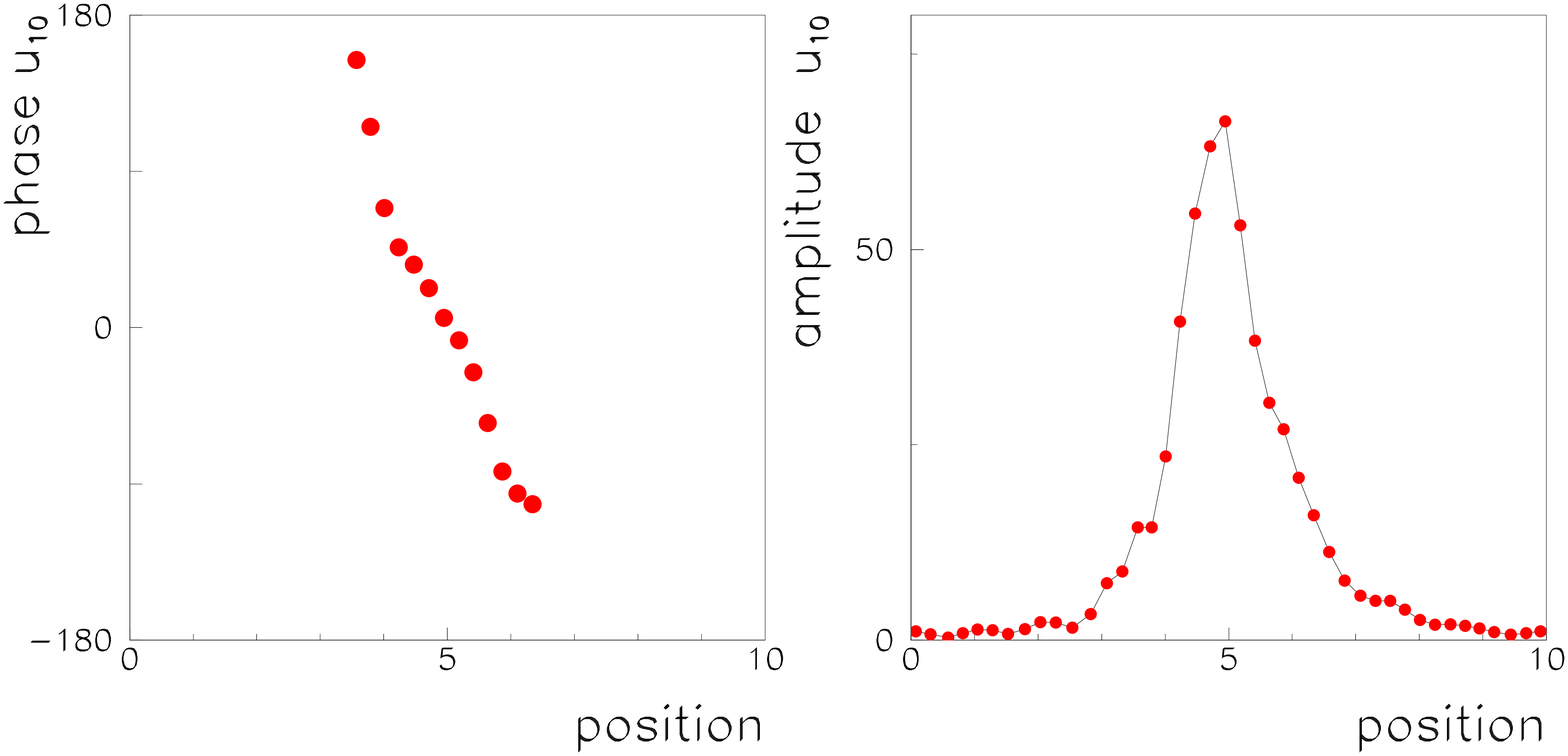}}
\put(8,0){C}
\end{picture}
\end{center}
\caption{
Phase precession persist after transient intra-hippocampal perturbation.
Despite the theta phase reset and the transient interruption of
firing, the phases were still correlated with the spatial
position of the animal immediately after the recovery,
in agreement with experiments (Zugaro et al 2005).%
A. Simulation of the network dynamics when
animal is moving continuously in time with constant 
velocity v=1/(200 ms) but a perturbation
that silences all units
(both excitatory and inhibitory ones)
and  reset the phase of theta rhythm is applied for 200 ms.
Black curves show  the theta rhythm. 
The dotted red line shows the
activity of the place cells
 before during and after
the network perturbation (place cell $u_8$, centered in location 4, in the upper
subplot, and place cell $u_10$, centered in 5, in the lower subplot).
The theta phase and the firing amplitude  are shown as a function of
 animal position when perturbation was applied, and
in control conditions (i.e. without any perturbation),
for the place cell centered in 4 (B) 
and the one centered in position 5 (C).
}
\label{ff3}
\end{figure*}


%
\begin{figure}[h]
\begin{center}
\includegraphics[width=8cm]{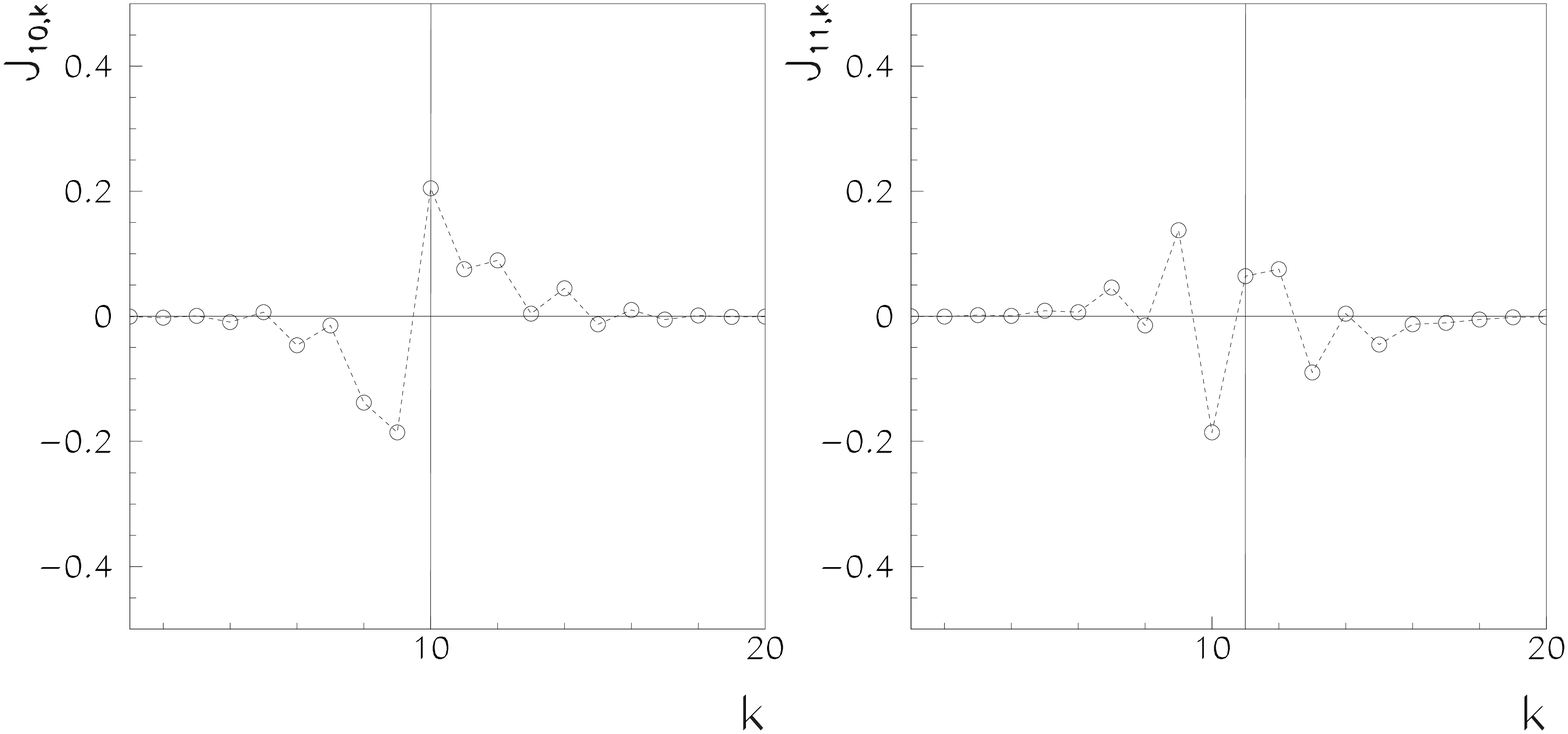}
\end{center}
\caption{%
$J_{ik}$ connection after learning N=10 locations, as a function of k.
On the left, $i = 10$, and on the right, $i = 11$.
}
\label{fj}
\end{figure}
\begin{figure}[h]
\begin{center}
\includegraphics[width=8cm]{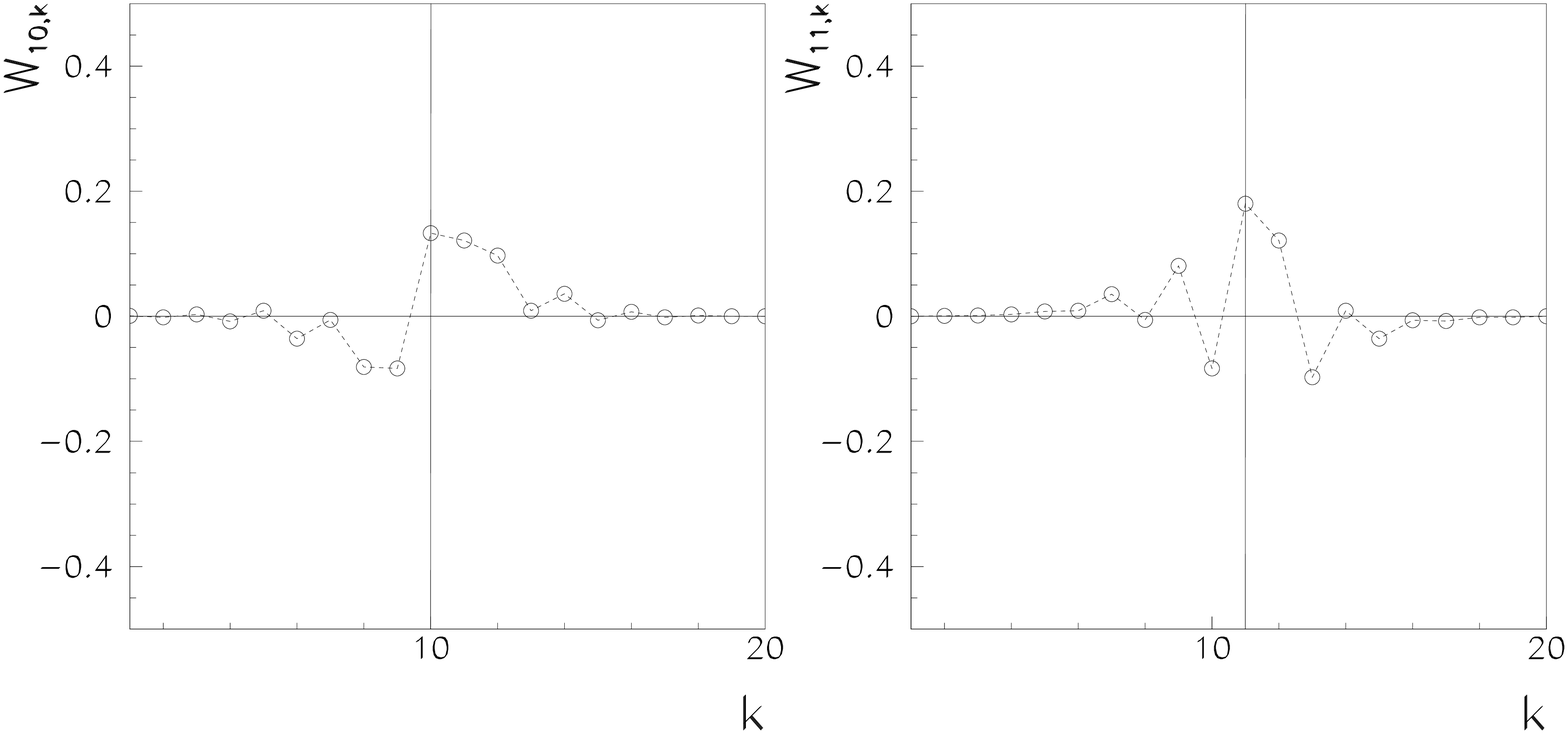}
\end{center}
\caption{%
$W_{ik}$ connection after learning N=10 locations, as a function of k.
On the left, $i = 10$, and on the right, $i = 11$.
}
\label{fw}
\end{figure}


\end{document}